\documentclass{article} 
\usepackage{nips15submit_e,times}
\usepackage{hyperref}
\usepackage{xcolor}
\usepackage{url}
\usepackage{amsmath,amsfonts}
\usepackage{amssymb}
\usepackage{graphicx,tabularx,array,geometry}
\usepackage[ruled,vlined]{algorithm2e}
\usepackage{etoolbox}
\usepackage{multirow}
\usepackage{subfigure}
\newtheorem{theorem}{Theorem}
\newtheorem{proposition}{Proposition}

\newlength\mylen
\newcommand\myinput[1]{%
  \settowidth\mylen{\KwIn{}}%
  \setlength\hangindent{\mylen}%
  \hspace*{\mylen}#1\\}
\renewcommand{\bar}{\overline}
\newcommand{\pa}{\partial}

\renewcommand{\Sigma}{\varSigma}

\newcommand{\vect}[1]{\boldsymbol{#1}}
\title{Hamiltonian Monte Carlo Acceleration using Surrogate Functions with Random Bases}

\author{
Cheng Zhang \\
Department of Mathematics\\
University of California, Irvine\\
Irvine, CA 92697 \\
\texttt{chengz4@uci.edu} \\
\And
Babak Shahbaba \\
Department of Statistics\\
University of California, Irvine\\
Irvine, CA 92697 \\
\texttt{babaks@uci.edu} \\
\AND
Hongkai Zhao\\
Department of Mathematics\\
University of California, Irvine\\
Irvine, CA 92697 \\
\texttt{zhao@math.uci.edu} \\
}

\nipsfinalcopy 

\begin{document}

\maketitle

\begin{abstract}
For big data analysis, high computational cost for Bayesian methods often limits their applications in practice. In recent years, there have been many attempts to improve computational efficiency of Bayesian inference. Here we propose an efficient and scalable computational technique for a state-of-the-art Markov Chain Monte Carlo (MCMC) methods, namely, Hamiltonian Monte Carlo (HMC). The key idea is to explore and exploit the structure and regularity in parameter space for the underlying probabilistic model to construct an effective approximation of its geometric properties. To this end, we build a surrogate function to approximate the target distribution using properly chosen random bases and an efficient optimization process. The resulting method provides a flexible, scalable, and efficient sampling algorithm, which converges to the correct target distribution. We show that by choosing the basis functions and optimization process differently, our method can be related to other approaches for the construction of surrogate functions such as generalized additive models or Gaussian process models. Experiments based on simulated and real data show that our approach leads to substantially more efficient sampling algorithms compared to existing state-of-the art methods.   
\end{abstract}

\section{Introduction}

Bayesian statistics has provided a principled and robust framework to create many important and powerful data analysis methods over the past several decades. Given a probabilistic model for the underlying mechanism of observed data, Bayesian methods properly quantify uncertainty and reveal the landscape or global structure of the parameter space. However, these methods tend to be computationally intensive since Bayesian inference usually requires the use of Markov Chain Monte Carlo (MCMC) algorithms to simulate samples from intractable distributions. Although the simple Metropolis algorithm \cite{metropolis53} is often effective at exploring low-dimensional distributions, it can be very inefficient for complex, high-dimensional distributions: successive states may exhibit high autocorrelation, due to the random walk nature of the movement. As a result, the effective sample size tends to be quite low and the convergence to the true distribution is usually very slow. The celebrated Hamiltonian Monte Carlo (HMC) \cite{duane87, neal11} reduces the random walk behavior of Metropolis by Hamiltonian dynamics, which uses gradient information to propose states that are distant from the current state, but nevertheless have a high probability of acceptance. 

Although HMC explores the parameter space more efficiently than random walk Metropolis does, it does not fully exploit the structure (i.e., geometric properties) of parameter space \cite{girolami11} since dynamics are defined over Euclidean space. To address this issue, Girolami and Calderhead \cite{girolami11} proposed a new method, called Riemannian Manifold HMC (RMHMC), that uses the Riemannian geometry of the parameter space \cite{amari00} to improve standard HMC's efficiency by automatically adapting to local structures. 

To make such geometrically motivated methods practical for big data analysis, one needs to combine them with efficient and scalable computational techniques. A common bottleneck for using such sampling algorithms for big data analysis is repetitive evaluations of functions, their derivatives, geometric and statistical quantities that involves the whole observed data and maybe a complicated model. A natural question is how to construct effective approximation of these quantities that provides a good balance between accuracy and computation cost. One common approach is subsampling (see, for example, \cite{wellingTeh11, hoffman10, shahbabaSplitHMC, chen14}), which restricts the computation to a subset of the observed data. This is based on the idea that big datasets contain a large amount of redundancy so the overall information can be retrieved from a small subset. However, in general applications, we cannot simply use random subsets for this purpose: the amount of information we lose as a result of random sampling leads to non-ignorable loss of accuracy, which in turn has a substantially negative impact on computational efficiency \cite{Betancourt15}. Therefore, in practice, it is a challenge to find good criteria and strategies for an appropriate and effective subsampling.

Another approach is to exploit smoothness or regularity in parameter space, which is true for most statistical models. This way, one could find computationally cheaper surrogate functions to substitute the expensive target (or potential energy) functions \cite{liu01, rasmussen03, zhang15, meeds14, lan15, strathmann15}. However, the usefulness of these methods is often limited to moderate dimensional problems because of the computational cost needed to achieve desired approximation accuracy.

In this work, our objective is to develop a faster alternative to the method of \cite{rasmussen03}. To this end, we propose to use random nonlinear bases along with efficient learning algorithms to construct a surrogate functions that provides effective approximation of the probabilistic model based on the collective behavior of the large data. The randomized nonlinear basis functions combined with the computationally efficient learning process can incorporate correct criteria for an efficient implicit subsampling resulting in both flexible and scalable approximation \cite{huang06,huang06chen,rahimi07,rahimi08}. Because our method can be presented as a special case of shallow random networks implemented in HMC, we refer to it as \emph{random network surrogate function}; however, we will show that our proposed method is related to (and can be extended to) other surrogate functions such as generalized additive models and Gaussian process models by constructing the surrogate functions using different bases and optimization process.

Our proposed method provides a natural framework to incorporate surrogate functions in the sampling algorithms such as HMC, and it can be easily extended to geometrically motivated methods such as Riemannian Manifold HMC. Further, for problems with a limited time budget, we propose an adaptive version of our method that substantially reduces the required number of training points. This way, the random bases surrogate function could be utilized earlier and its approximation accuracy could be improved adaptively as more training points become available. We show that theoretically the learning procedure for our surrogate function is asymptotically equivalent to {\em potential matching}, which is itself a novel distribution matching strategy similar to the {\em score matching} method discussed in \cite{hyvarinen05, strathmann15}. 

Finally, we should emphasize that out method is used to generate high quality proposals at low computational cost. However, when calculating the acceptance probability of these proposals, we use the original Hamiltonian (used in standard HMC) to ensure that the stationary distribution of the Markov chain will remain the correct target distribution. 

Our paper is organized as follows. An overview of HMC and RMHMC is given in Section \ref{sec:preliminaries}. Our random network surrogate HMC is explained in detail in Section \ref{sec:RNSHMC}. The adaptive model is developed in Section \ref{sec:OLRNSHMC}. Experimental results based on simulated and real data are presented in Section \ref{sec:results}. Code for these examples is available at \textcolor{blue}{\url{github.com/chengzhang-uci/RNSHMC}.} Finally, Section \ref{sec:discussion} is devoted to discussion and future work. As an interesting supplement, an overall description of the {\em potential matching} procedure is presented in Section \ref{sec:potentialmatching} in the appendix. 

\section{Preliminaries}
\label{sec:preliminaries}
\subsection{Hamiltonian Monte Carlo}
In Bayesian Statistics, we are interested in sampling from the posterior distribution of the model parameters $q$ given the observed data, $Y=(y_1,y_2,\ldots,y_N)^T$,
\begin{equation}
P(q|Y) \propto \exp(-U(q)),
\end{equation}
where the potential energy function $U$ is defined as
\begin{equation} \label{eq:U}
U(q) = -\sum_{i=1}^N\log P(y_i|q) -\log P(q).
\end{equation}
The posterior distribution is almost always analytically intractable. Therefore, MCMC algorithms are typically used for sampling from the posterior distribution to perform statistical inference. As the number of parameters increases, however, simple methods such as random walk Metropolis \cite{metropolis53} and Gibbs sampling \cite{geman84} may require a long time to converge to the target distribution. Moreover, their explorations of  parameter space become slow due to inefficient random walk proposal-generating mechanisms, especially when there exist strong dependencies between parameters in the target distribution. By inclusion of geometric information from the target distribution, HMC \cite{duane87,neal11} introduces a Hamiltonian dynamics system with auxiliary momentum variables $p$ to propose samples of $q$ in a Metropolis framework that explores the parameter space more efficiently compared to standard random walk proposals. More specifically, HMC generates proposals jointly for $q$ and $p$ using the following system of differential equations:
\begin{align}
\frac{dq_i}{dt} &= \frac{\partial H}{\partial p_i}\label{eq:Hq}\\
\frac{dp_i}{dt} &= -\frac{\partial H}{\partial q_i}\label{eq:Hp}
\end{align}
where the Hamiltonian function is defined as $H(q,p) = U(q) + \frac12p^TM^{-1}p$. The quadratic kinetic energy function $K(p) = \frac12p^TM^{-1}p$ corresponds to the negative log-density of a zero-mean multivariate Gaussian distribution with the covariance $M$. Here, $M$ is known as the mass matrix, which is often set to the identity matrix, $I$. Starting from the current state $(q,p)$, the Hamiltonian dynamics system \eqref{eq:Hq},\eqref{eq:Hp} is simulated for $L$ steps using the leapfrog method, with a stepsize of $\epsilon$. The proposed state, $(q^{\ast},p^{\ast})$, which is at the end of the trajectory, is accepted with probability $\min(1,\exp[-H(q^{\ast},p^{\ast}) + H(q,p)])$. By simulating the Hamiltonian dynamics system together with the correction step, HMC generates samples from a joint distribution
\[
P(q,p) \propto \exp\left(-U(q)-\frac12p^TM^{-1}p\right)
\]
Notice that $q$ and $p$ are separated, the marginal distribution of $q$ then follows the target distribution.
These steps are presented in Algorithm \ref{alg:HMC}.
\begin{algorithm}[t]
{\small
\KwIn{Starting position $q^{(1)}$ and step size $\epsilon$}
 \For{$t =1,2,\cdots$ }{
  \textit{Resample momentum $p$}\\
  $p^{(t)} \sim \mathcal{N}(0,M),\;(q_0,p_0)$ = $(q^{(t)},p^{(t)})$\\
  \textit{Simulate discretization of Hamiltonian dynamics:}\\
  \For{$l = 1$ to $L$} {
  $p_{l-1} \leftarrow p_{l-1} - \frac{\epsilon}{2} \frac{\partial U}{\partial q}(q_{l-1})$\\
  $q_l \leftarrow q_{l-1} + \epsilon M^{-1}p_{l-1}$\\
   $p_l \leftarrow p_l - \frac{\epsilon}{2} \frac{\partial U}{\partial q}(q_{l})$
  }
  $(q^{\ast},p^{\ast}) = (q_L,p_L)$\\
  \textit{Metropolis-Hasting correction:} \\
  $u \sim \text{Uniform}[0,1]$\\
  $\rho = \exp[{ H(q^{(t)},p^{(t)})-H(q^{\ast},p^{\ast}) }]$\\
  \lIf{$u < \min(1,\rho)$,} {$q^{(t+1)} = q^{\ast}$}
 }
\caption{Hamiltonian Monte Carlo}
}
\label{alg:HMC}
\end{algorithm}
Following the dynamics of the assumed Hamiltonian system, HMC can generate distant proposals with high acceptance probability which allows an efficient exploration of parameter space. 

\subsection{Riemannian Manifold HMC}
Although HMC explores the target distribution more efficiently than random walk Metropolis, it does not fully exploit the geometric structures of the underlying probabilistic model since a flat metric (i.e., ${M} = {I}$) is used. Using more geometrically motivated methods could substantially improves sampling algorithms' efficiency. Recently, Girolami and Calderhead \cite{girolami11} proposed a new method, called Riemannian Manifold HMC (RMHMC), that exploits the Riemannian geometry of the target distribution to improve standard HMC's efficiency by automatically adapting to local structures. To this end, instead of the identity mass matrix commonly used in standard HMC, they use a position-specific mass matrix ${M} = { G}({q})$. More specifically, they set ${G}({q})$ to the Fisher information matrix, and define Hamiltonian as follows:
\begin{equation}\label{rmhamiltonp}
H({q}, { p})  = U({q}) +\frac12 \log\det { G}({q}) + 
\frac12 { p}^{T}{ G}({q})^{-1}{ p} = \phi(q) + \frac12 { p}^{T}{ G}({q})^{-1}{ p}
\end{equation}
where $\phi(q):= U(q) +\frac12 \log\det G(q)$. Note that standard HMC is a special case of RMHMC with $G(q) =  I$. Based on this dynamic, they propose the following HMC on Riemmanian manifold:
\begin{eqnarray}\begin{array}{lcrcr }
\displaystyle 
\dot q & = & \nabla_p H(q, p) & = & G(q)^{-1}p \\ [12pt]
\displaystyle 
\dot p & = & -\nabla_q H(q, p)& =  & -\nabla_q \phi(q) + 
\frac12 \nu(q,p) 
\end{array}\label{eq:rmhmc}\end{eqnarray}
where the $i$th element of the vector $\nu(q,p)$ is 
\begin{eqnarray*}
(\nu(q,p))_i = -p^{T} \pa_i ( G(q)^{-1})p = ({G(q)^{-1}p})^{T} \pa_i G(q) {G(q)^{-1}p}
\end{eqnarray*}
with the shorthand notation $\partial_i =  {\partial}/{\partial q_i}$ for partial derivative.

The above dynamic is non-separable (it contains products of $q$ and $p$), and the resulting proposal generating mechanism based on the standard leapfrog method is neither time-reversible nor symplectic. Therefore, the standard leapfrog algorithm cannot be used for the above dynamic \cite{girolami11}. Instead, we can use the St\"omer-Verlet \cite{verlet67} method, known as generalized leapfrog \cite{leimkuhler04},
\begin{eqnarray}\label{gleapfrog}
p^{(t+1/2)} 
 & = & p^{(t)} - \frac{\epsilon}{2} \left[\nabla_q\phi(q^{(t)})-\frac12
\nu(q^{(t)},p^{(t+1/2)})\right] \label{gleapfrog:imp1}\\
q^{(t+1)} 
 & = & q^{(t)} + \frac{\epsilon}{2} \left[G^{-1}(q^{(t)}) + G^{-1}(q^{(t+1)})\right]p^{(t+1/2)}\label{gleapfrog:imp2}\\
p^{(t+1)} 
 & = & p^{(t+1/2)} - \frac{\epsilon}{2}\left[\nabla_q\phi(q^{(t+1)})-\frac12
\nu(q^{(t+1)},p^{(t+1/2)})\right]
\end{eqnarray}
The resulting map is 1) deterministic, 2) reversible, and 3) volume-preserving. However, it requires solving two computationally intensive implicit equations (Equations \eqref{gleapfrog:imp1} and \eqref{gleapfrog:imp2}) at each leapfrog step.

\section{Random Network Surrogate HMC (RNS-HMC)}
\label{sec:RNSHMC}
For HMC, the Hamiltonian dynamics contains the information from the target distribution through the potential energy $U$ and its gradient. For RMHMC, more geometric structure (i.e., the Fisher information) is included through the mass matrix for kinetic energy. It is the inclusion of such information in the Hamiltonian dynamics that allows HMC and RMHMC to improve upon random walk Metropolis. However, one common computational bottleneck for HMC and other Bayesian models for big data is repetitive evaluations of functions, their derivatives, geometric and statistical quantities. Typically, each evaluation involves the whole observed data. For example, one has to compute the potential $U$ and its gradient from the equation (\ref{eq:U}) for HMC and mass matrix $M$, its inverse and the involved partial derivatives for RMHMC \emph{at every time step or M-H correction step}. When $N$ is large, this can be extremely expensive to compute. In some problems, each evaluation may involve solving a computationally expensive problem. (See the inverse problem and Remark in Section \ref{sec:PDE}.)

To alleviate this issue, in recent years several methods have been proposed to construct \emph{surrogate} Hamiltonians. For relatively low dimensional spaces, (sparse) grid based piecewise interpolative approximation using precomputing strategy was developed in \cite{zhang15}. Such grid based methods, however, are difficult to extend to high dimensional spaces due to the use of structured grids. Alternatively, we can use Gaussian process model, which are commonly used as surrogate models for emulating expensive-to-evaluate functions, to learn the target functions from early evaluations (training data) \cite{rasmussen03, lan15, meeds14}. However, naive (but commonly used) implementations of Gaussian process models have high computation cost associated with inverting the covariance matrix, which grows cubically as the size of the training set increases. This is especially crucial in high dimensional spaces, where we need large training sets in order to achieve a reasonable level of accuracy. Recently, scalable Gaussian processes using induing point methods \cite{snelson06,joaquin05} have been introduced to scale up GPs to larger datasets. While these methods have been quite succsessful in reducing computation cost, the tuning of inducing points could still be problematic in high dimensional spaces. (See a more detailed discussion in Section \ref{sec:SPGP}.)

The key in developing surrogate functions is to develop a method that can effectively capture the collective properties of large datasets with scalability, flexibility and efficiency. In this work, we propose to construct surrogate functions using proper random nonlinear bases and efficient optimization process on training data. More specifically, we present our method as a special case of shallow neural networks; although, we show that it is related to (and can be extended to) other surrogate functions such as generalized additive models and Gaussian process models. Random networks of nonlinear functions prove capable of approximating a rich class of functions arbitrarily well \cite{huang06chen,rahimi08}. Using random nonlinear networks and algebraic learning algorithms can also be viewed as an effective implicit subsampling with desired criteria. The choice of hidden units (basis functions) and the fast learning process can be easily adapted to be problem specific and scalable. Unlike typical (naive) Gaussian process models, our random network scales linearly with the number of training points. In fact, a random nonlinear network can be considered as a standard regression model with randomly mapped features. For such shallow random networks, the computational cost for inference is cubic in the number of hidden nodes. Those differences in scaling allow us to explicitly trade off computational efficiency and approximation accuracy and construct more efficient surrogate in certain applications. As our empirical results suggest, with appropriate training data good approximation of smooth functions in high dimensional space can be achieved using a moderate and scalable number of hidden units. Therefore, our proposed method has the potential to scale up to large data sizes and provide effective and scalable surrogate Hamiltonians that balance accuracy and efficiency well. 


\subsection{Shallow Random Network Approximation}
A typical shallow network architecture (i.e., a single-hidden layer feedforward scalar-output neural network) with $s$ hidden units, a nonlinear activation function $a$, and a scalar (for simplicity) output $z$ for a given {\it d}-dimensional input $q$ is defined as 
\begin{equation}\label{eq:nn}
z(q) = \sum_{i=1}^sv_ia(q;\gamma_i)+b
\end{equation}
where $\gamma_i$ is the {\it i}th hidden node parameter, $v_i$ is the output weight for the {\it i}th hidden node, and $b$ is the output bias. Given a training data set 
\[\mathcal{T} = \{(q^{(j)},t^{(j)})|q^{(j)}\in\mathbb{R}^d,\;t^{(j)}\in\mathbb{R},\; j = 1,\ldots,N\}\] the neural network can be trained by finding the optimal model parameters $\vect{W}=\{\gamma_i,v_i,\;i=1,\ldots,s,b\}$ to minimize the mean square error cost function,
\begin{equation}\label{eq:cost}
C(\vect{W}|\mathcal{T}) = \frac1N\sum_{j=1}^N\|z(q^{(j)})-t^{(j)}\|^2
\end{equation}

The most popular algorithm in machine learning to optimize \eqref{eq:cost} is back-propagation \cite{rumelhart86}. However, as a gradient descent-based iterative algorithm, back-propagation is usually quite slow and can be trapped at some local minimum since the cost function is nonlinear, and for most cases, non-convex. Motivated by the fact that randomization is computationally cheaper than optimization, alternative methods based on random nonlinear bases have been proposed \cite{huang06,rahimi07}. These methods drastically decrease the computation cost while maintaining a reasonable level of approximation accuracy. The key feature of random networks is that they reduce the full optimization problem into standard linear regression by mapping the input data to a randomized feature space and then apply existing fast algebraic training methods (e.g., by minimizing squared error) to find the output weight.  Given the design objective, algebraic training can achieve exact or approximate matching of the data at the training points. Compared to the gradient descent-based techniques, algebraic training methods can reduce computational complexity and provide better generalization properties. A typical algebraic approach for single-hidden layer feedforward random networks is extreme learning machine (ELM) \cite{huang06}, which is summarized in Algorithm \ref{alg:ELM}.
\begin{algorithm}[t]
{\small
\caption{Extreme Learning Machine}\label{alg:ELM}
Given a training set $\mathcal{T} = \{(I_j,t_j)|I_j\in\mathbb{R}^d,\;t_j\in\mathbb{R}^m,\; j = 1,\ldots,N\}$, activation function $a(x;\gamma)$ and hidden node number $s$\\[3pt]
Step 1: Randomly assign hidden node parameters $\gamma_i$, $i = 1,\ldots,s$ \\[3pt]
Step 2: Calculate the hidden layer output matrix $H$
\[
H_{ji} = a(I_j;\gamma_i),\quad i=1,\ldots,s,\;j = 1,\ldots,N
\]
Step 3: Calculate the output weight $v$
\[
v = H^{\dagger}T,\quad T =[t_1,t_2,\ldots,t_N]^T
\]
\hspace{30pt}where $H^{\dagger}$ is the \textit{Moore-Penrose generalized inverse} of matrix $H$
}
\end{algorithm}
Using randomized nonlinear features, ELM estimates the output weight by finding the least-squares solution to the resulting linear equations system $Hv = T$. Note that presented this way, our method can also be regarded as a random version of Generalized Additive Model (GAM). In practice, people could add regularization to improve stability and generalizability.

\subsection{Choice of Nonlinearity}
There are many choices for nonlinear activation functions in random networks. Different types of activation functions can be used for different learning tasks. In this paper, we focus on random networks with two typical types of nonlinear nodes: 
\begin{itemize}
\item \emph{Additive nodes}:
\[
a(q;\gamma) = a(\vect{w}\cdot q + d),\quad \vect{w}\in\mathbb{R}^d,\;d \in\mathbb{R}, \quad \gamma = \{\vect{w},d\}
\]
where $\vect{w}$ and $d$ are the weight vector and the bias of the hidden node.
\item \emph{Radial basis functions (RBF) nodes}:
\[
a(q;\gamma) = a\left(-\frac{\|q-c\|^2}{2\ell^2}\right),\quad  c\in\mathbb{R}^d,\;\ell \in\mathbb{R}^{+}, \quad \gamma = \{c,\ell\}
\]
where $c$ and $\ell$ are the center and width of the hidden node.
\end{itemize}

Both random networks can approximate a rich class of functions arbitrarily well \cite{huang06chen,rahimi08}. With randomly assigned input weights and biases composed linearly inside the nonlinear activation function, additive nodes form a set of basis functions, whose level sets are hyperplanes orientated by $\vect{w_i}$  and shifted by $d_i$ respectively. Random networks with additive nodes tend to reflect the global structure of the target function. On the other hand, RBF nodes are almost compactly supported (can be adjusted by the width $\ell$) rendering good local approximation for the corresponding random networks.

\subsection{Connection to GPs and Sparse GPs}\label{sec:SPGP}
It is worth noting the connection between networks with RBF nodes and Gaussian processes models \cite{rasmussen96,neal99}. Given a training data set 
\[\mathcal{T} = \{(q^{(j)},t^{(j)})|q^{(j)}\in\mathbb{R}^d,\;t^{(j)}\in\mathbb{R},\; j = 1,\ldots,N\}\]
and using squared exponential covariance function
 \[
 K(q^{(j)},q^{(j')}) = \sigma_f^2\exp\left(-\frac{\|q^{(j)}-q^{(j')}\|^2}{2\ell^2}\right),\quad \theta = \{\sigma_f,\ell\}
 \] 
the standard GP regression with a Gaussian noise model has the following marginal likelihood
\[
p(\vect{t}|\vect{Q},\theta) = \mathcal{N}(\vect{t}|\vect{0},\vect{K}_N+\sigma^2\vect{I}) 
\]
where $\vect{Q}=\{q^{(j)}\}_{j=1}^N, \; \vect{t} = \{t^{(j)}\}_{j=1}^N,\; [\vect{K}_N]_{jj'} = K(q^{(j)},q^{(j')})$ is the covariance matrix and $\sigma$ is the noise parameter. Prediction on a new observation $q^{\ast}$ is made according to the conditional distribution
\[
p(t^{\ast}|q^{\ast},\mathcal{T},\theta) = \mathcal{N}(t^{\ast}|\vect{k}_\ast^T(\vect{K}_N+\sigma^2\vect{I})^{-1}\vect{t},\;K_{\ast\ast}-\vect{k}_\ast^T(\vect{K}_N+\sigma^2\vect{I})^{-1}\vect{k}_\ast + \sigma^2)
\]
where $[k_\ast]_j=K(q^{(j)},q^{\ast})$ and $K_{\ast\ast} = K(q^{\ast},q^{\ast})$.

On the other hand, if we use $K(q^{(j)},\cdot)$ as the $j$th hidden node,  $j=1,2,\ldots,N$, the output matrix becomes $H=\vect{K}_N$, and the output weight learned by algebraic approach to a regularized least square problem is 
\[
\hat{v} = \arg\min_{v} \|\vect{H}v-\vect{t}\|^2 + \sigma^2v^T\vect{K}_Nv =  (\vect{K}_N+\sigma^2\vect{I})^{-1}\vect{t}
\]
Therefore, such a network provides the same prediction point estimate as the above full GP models. This way, a Gaussian process model can be interpreted as a self-organizing RBF network where new hidden nodes are added adaptively with new observations. This is also an alternative point of view to \cite{neal96} where GP models were shown to be equivalent to single hidden-layer neural networks with infinite many hidden nodes.

Notice that the above GP model scales cubically with the number of data points $N$ which limits the application of GPs to relatively small datasets. To derive a GP model that is computationally tractable for large datasets, sparse GPs based on inducing point methods \cite{snelson06,joaquin05} have been previously proposed. These sparse models introduce a set of inducing points $\bar{\vect{Q}}= \{\bar{q}^{(i)}\}_{i=1}^M$ and approximate the exact kernel $K(q,q')$ by an approximation $\tilde{K}(q,q')$ for fast computation. For example, the fully independent training conditional (FITC) \cite{snelson06} method uses the approximate kernel
\[
\tilde{K}_{\small{\mathrm{FITC}}}(q,q')= \vect{k}_q^T\vect{K}_M^{-1}\vect{k}_{q'} + \delta_{qq'}(K(q,q')-\vect{k}_q^T\vect{K}_M^{-1}\vect{k}_{q'})
\]
where $\vect{K}_M$ is the exact covariance matrix for the $M$ inducing points and $\vect{k}_q,\vect{k}_{q'}$ are the exact covariance matrices between $q,q'$ and the inducing points. Given the same training data set $\mathcal{T}$, the marginal likelihood is
\[
p(\vect{t}|\vect{Q},\theta) = \mathcal{N}(\vect{t}|\vect{0},\vect{K}_{NM}\vect{K}_M^{-1}\vect{K}_{MN}+\vect{\Lambda}+\sigma^2\vect{I})
\]
Here, $\vect{\Lambda}$ is a diagonal matrix with $\vect{\Lambda}_{jj} = K_{jj}- \vect{k}_j^T\vect{K}_{M}^{-1}\vect{k}_j$ that adjusts the diagonal elements to the ones from the exact covariance matrix $\vect{K}_N$. Simiarly, predictions can be made according to the conditional distribution
\[
p(t^{\ast}|q^{\ast},\mathcal{T},\theta) = \mathcal{N}(t^{\ast}|\vect{k}_\ast^T\vect{\Sigma}_M^{-1}\vect{K}_{MN}(\vect{\Lambda}+\sigma^2\vect{I})^{-1}\vect{t},K_{\ast\ast}-\vect{k}_\ast^T(\vect{K}_M^{-1}-\vect{\Sigma}_M^{-1})\vect{k}_\ast + \sigma^2)
\]
where $\vect{\Sigma}_M = \vect{K}_M + \vect{K}_{MN}(\vect{\Lambda}+\sigma^2\vect{I})^{-1}\vect{K}_{NM}$. The computation cost is reduced to $\mathcal{O}(M^2N)$ for learning and after that the predictive mean costs $\mathcal{O}(M)$ per new observation. The hyperparameters $\theta,\sigma$ and inducing points $\bar{\vect{Q}}$ can be tuned to maximize the marginal likelihood. However, in high dimension the tuning of $\bar{\vect{Q}}$ becomes infeasible.

\begin{figure}[!t]
\begin{center}
\includegraphics[width=0.33\textwidth,height=0.3\textheight]{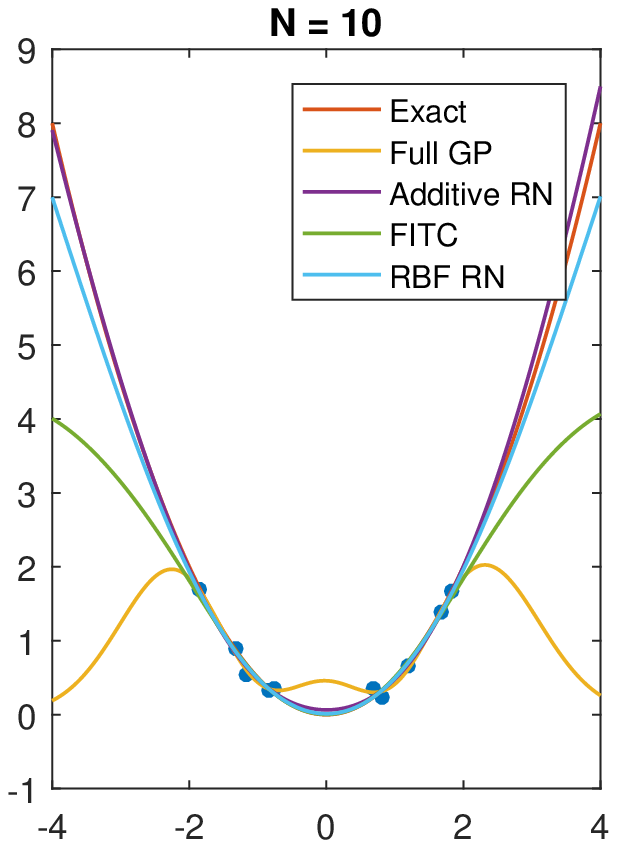}  \hspace{-8pt}
\includegraphics[width=0.33\textwidth,height=0.3\textheight]{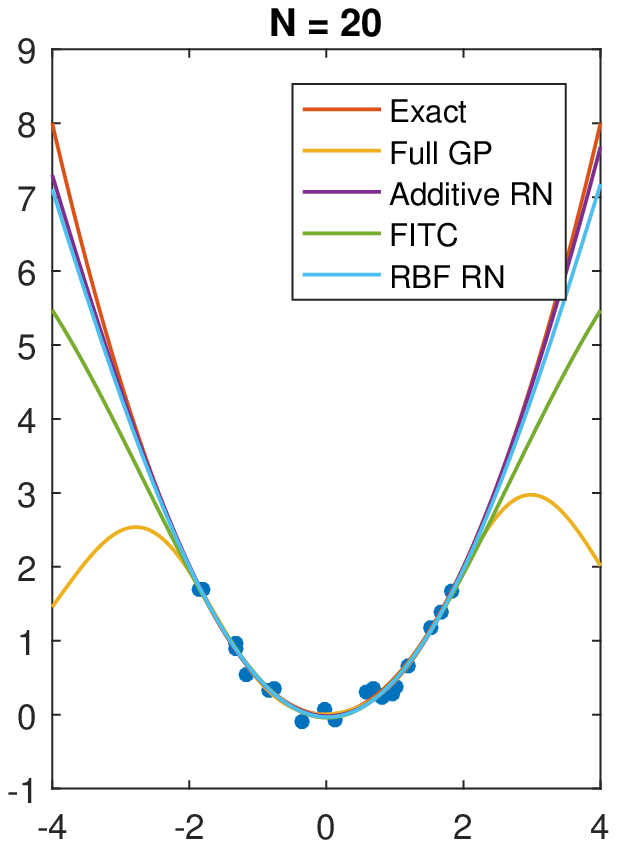} \hspace{-8pt}
\includegraphics[width=0.33\textwidth,height=0.3\textheight]{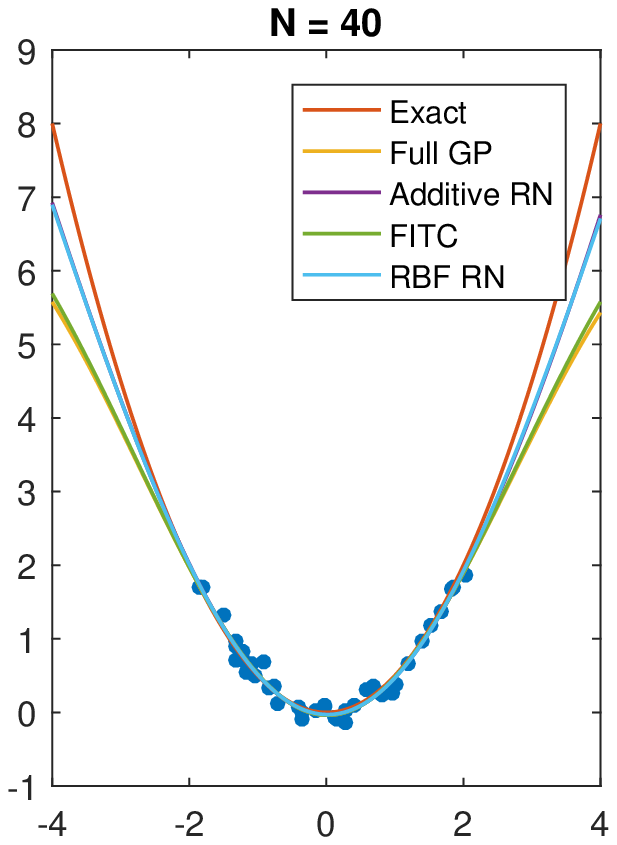}   \hspace{-8pt}\\
\caption{Comparing different surrogate approximations with an increasing number of observations $N=10,20,40$ on target function $y=x^2/2$. The observation points are nested samples from the standard normal distribution. For FITC and random networks, we choose 5 inducing points and 5 hidden neurons respectively. FITC and random networks are all run 100 times and averaged to reduce the random effects on the results.}\label{fig:surrcomp}
\vspace{-10pt}
\end{center}
\end{figure}

On the other hand, if we use the inducing points $\bar{\vect{Q}}$ as the centers of a RBF network, the output matrix $\vect{H}=\vect{K}_{NM}$. Given the diagonal matrix $\vect{D} = \vect{\Lambda} + \sigma^2\vect{I}$, the output weight estimated by the algebraic approach to a weighted least square problem plus a regularization term is 
\[
\hat{v} = \arg\min_v \|\vect{D}^{-\frac12}(\vect{H}v-\vect{t})\|^2 + v^T\vect{K}_Mv = \vect{\Sigma}_M^{-1}\vect{K}_{MN}(\vect{\Lambda}+\sigma^2\vect{I})^{-1}\vect{t}
\]
Therefore, the same predictive mean can be reproduced if we use the inducing points as centers and use the same hyperparameter configuration in our random network with RBF nodes and optimize with respect to the above cost function. Moreover, those hyperparameters in each basis (or hidden nodes)  can also be trained (typically by gradient decent) if we abandon the use of random bases and simple linear regression. 

Figure \ref{fig:surrcomp} compares random networks with different node types and related GP methods on fitting a simple function $y=x^2/2$ which corresponds to the negative log-density of the standard normal distributions. We used \emph{softplus} function $\sigma(x) = \log\big(1+\exp(x)\big)$ in additive nodes and exponential square kernels in RBF nodes and GP methods. As we can see from the graph, random networks generally perform better than GP methods when the number of observations is small. The randomness in the configuration of hidden nodes force networks to learn more globally. In contrast, GP models are more local and need more data to generalize well. By introducing sparsity, FITC tends to generalize better than full GP, especially on small datasizes. Since our goal is to fit negative log-posterior density function in (\ref{eq:U}) and $U(q)\rightarrow \infty$ as $q$ moves away from the high density domain, using \emph{softplus} basis functions in random networks are more capable to capture this far field feature by striking a better balance between flexibility and generalization while being less demanding on the datasize. Also, the number of hidden neurons (bases) can be used to regularize the approximation to mitigate overfitting issue.

\subsection{Surrogate Induced Hamiltonian Flow}
\begin{figure}[!t]
\begin{center}
\includegraphics[width=0.33\textwidth,height=0.3\textheight]{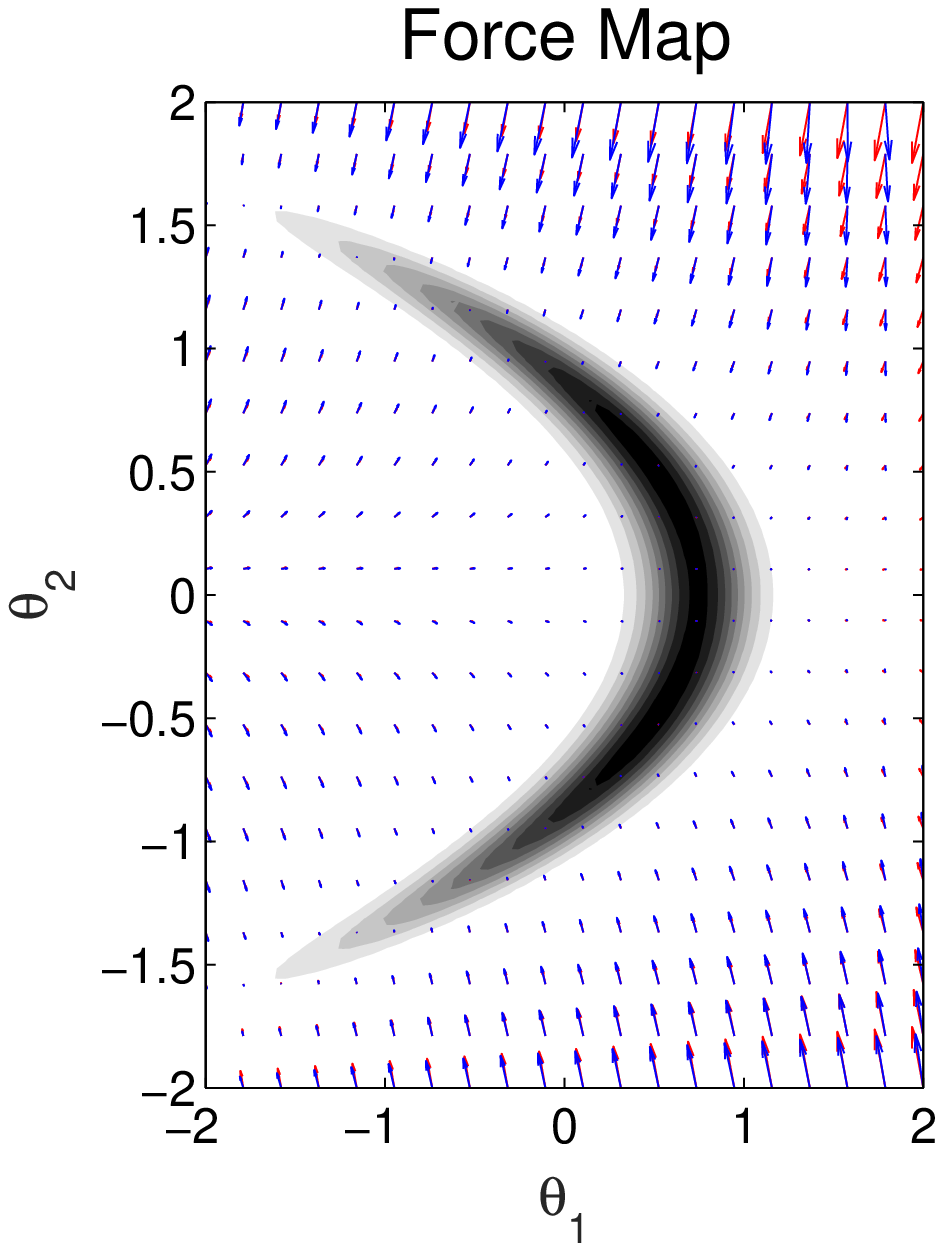}  \hspace{-8pt}
\includegraphics[width=0.33\textwidth,height=0.3\textheight]{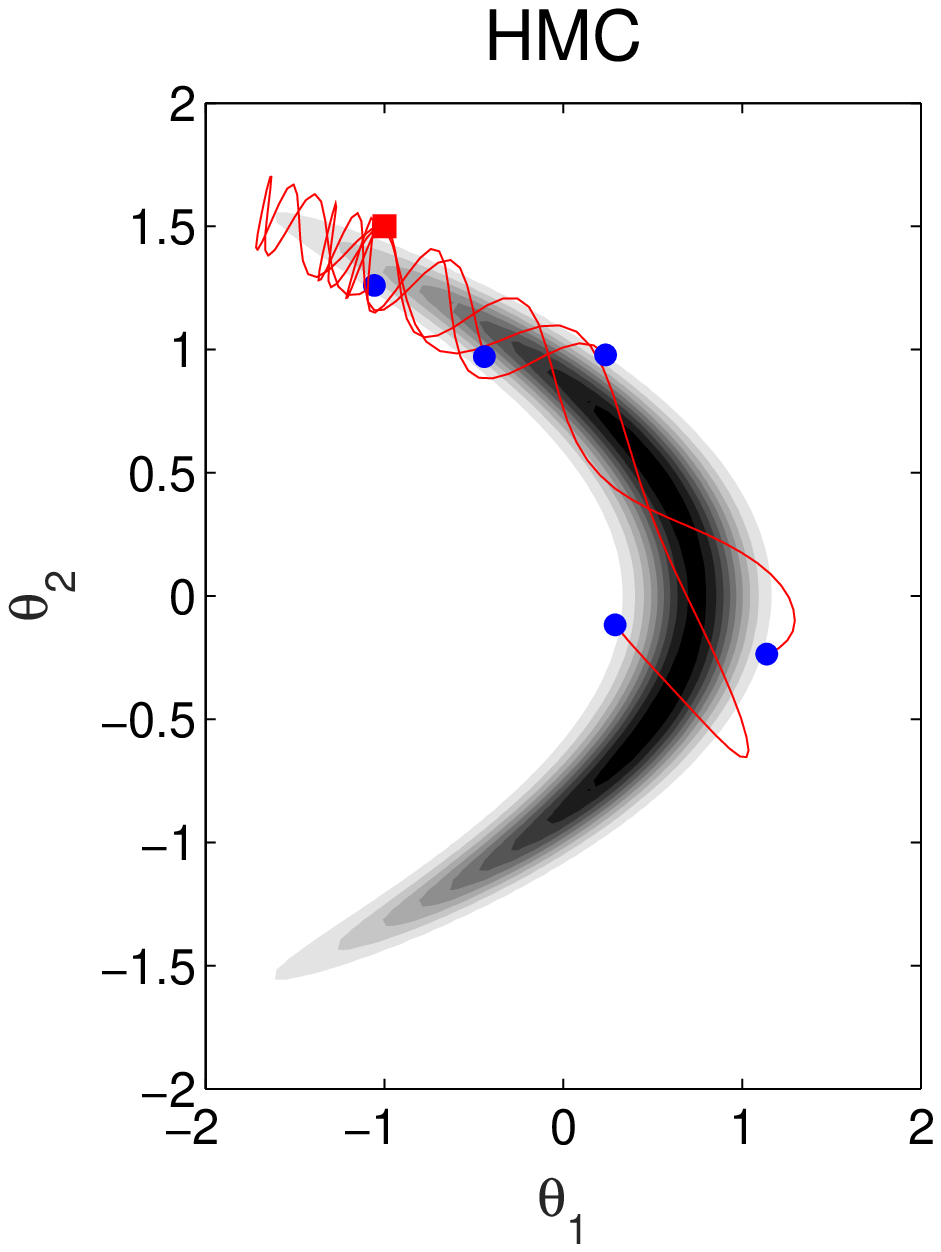} \hspace{-8pt}
\includegraphics[width=0.33\textwidth,height=0.3\textheight]{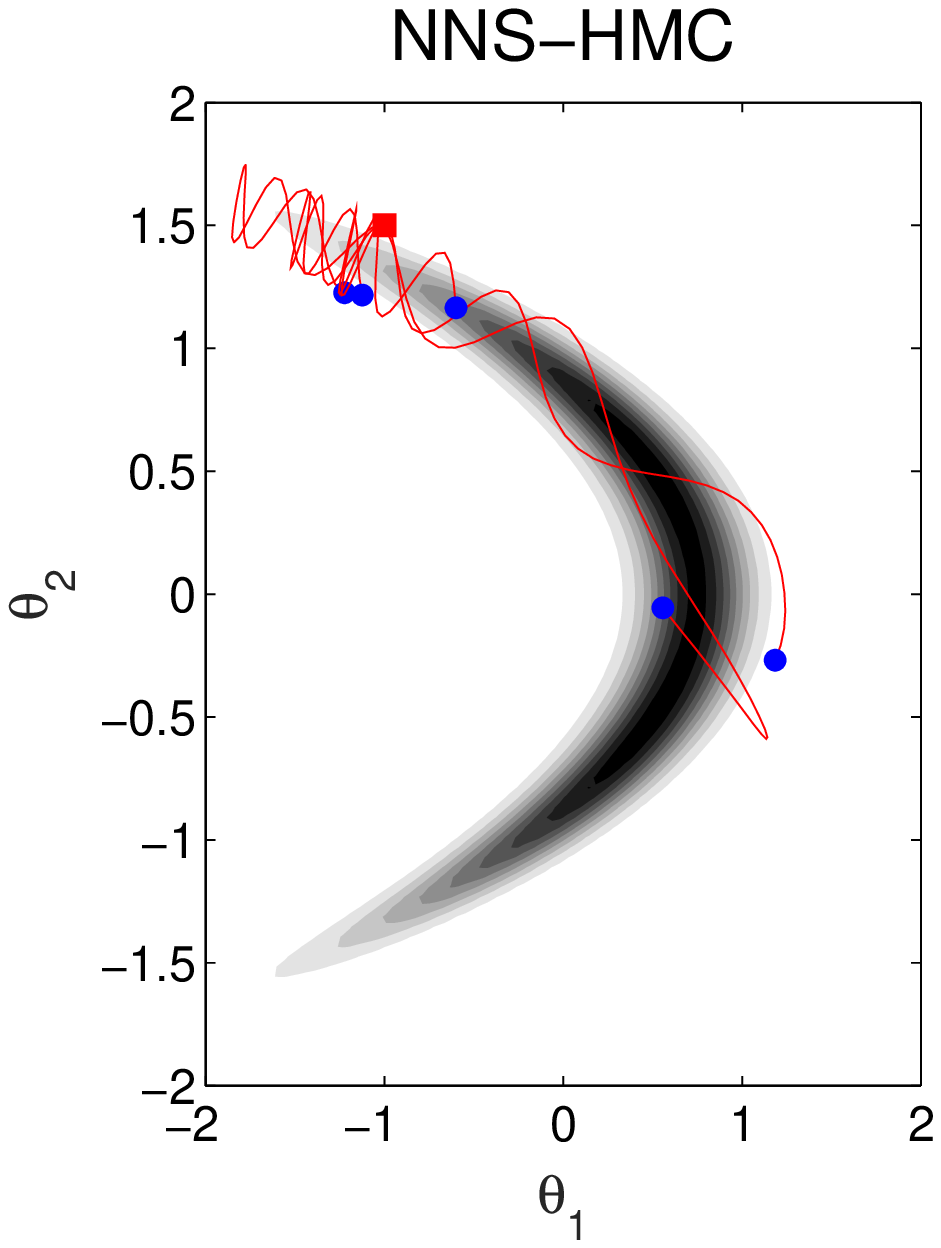}   \hspace{-8pt}
\caption{Comparing HMC and NNS-HMC based on a 2-dimensional banana-shaped distribution. The left panel shows the gradient fields (force map) for the original Hamiltonian flow (red) and the surrogate induced Hamiltonian flow (blue). The middle and right panel show the trajectories for HMC and NNS-HMC samplers. Both samplers start from the same point (red square) with same initial momentums. Blue points at the end of the trajectories are the proposals. The overall acceptance probability drops from 0.97 using HMC to 0.88 using NNS-HMC.}\label{fig:ar}
\vspace{-10pt}
\end{center}
\end{figure} 

As mentioned in the previous sections, repetitive computation of Hamiltonian, its gradient and other quantities that involve all data set undermine the overall exploration efficiency of HMC. To alleviate this issue, we exploit the smoothness or regularity in parameter space, which is true for most statistical models. As discussed in \cite{neal96, liu01, rasmussen03}, one can improve computational efficiency of HMC by approximating the energy function and using the resulting approximation to device a surrogate transition mechanism while still converging to the correct target distribution. To this end, \cite{rasmussen03} proposed to use pre-convergence samples (which are discarded during the burn-in period) to approximate the energy function using a Gaussian process model. Here, we define an alternative surrogate-induced Hamiltonian as follows:
\[
\tilde{H}(q,p) = \tilde{U}(q) + \frac12p^TM^{-1}p
\]
where $\tilde{U}(q)$ is the neural network surrogate function. $\tilde{H}(q,p)$ now defines a surrogate-induced Hamiltonian flow, parametrized by a trajectory length $t$, which is a map $\tilde{\phi}_t:\; (q,p) \rightarrow (q^{\ast},p^{\ast})$. Here, $(q^{\ast},p^{\ast})$ being the end-point of the trajectory governed by the following equations
\[
\frac{dq}{dt} = \frac{\partial \tilde{H}}{\partial p} = M^{-1}p,\quad \frac{dp}{dt} = - \frac{\partial \tilde{H}}{\partial q} = -\frac{\partial \tilde{U}}{\partial q}
\]
When the original potential $U(q)$ is computationally costly,  simulating the surrogate induced Hamiltonian system provides a more efficient proposing mechanism for our HMC sampler. The introduced bias along with discretization error from the leap-frog integrator are all naturally corrected in the MH step where we use the original Hamiltonian in the computation of acceptance probability. As a result, \emph{the stationary distribution of the Markov chain will remain the correct target distribution} (see Appendix \ref{sec:target} for a detailed proof). Note that by controlling the approximation quality of the surrogate function, we can maintain a relatively high acceptance probability. This is illustrated in Figure \ref{fig:ar} for a two-dimensional banana-shaped distribution \cite{girolami11}.


To construct such a surrogate, the early evaluations of the target function during the early iterations of MCMC will be used as the training set based on which we can train a shallow random network using fast algebraic approaches, such as ELM (Algorithm \ref{alg:ELM}). The gradient of the scalar output $z$ (see \ref{eq:nn}) for a network with additive hidden nodes, for example, then can be computed as
\begin{equation}\label{eq:grad}
\frac{\partial z}{\partial q} = \sum_{i=1}^sv_ia'(\vect{w}_i\cdot q+d_i)\vect{w}_i
\end{equation}
which costs only $\mathcal{O}(s)$ computations. To balance the efficiency in computation and flexibility in approximation, and to reduce the possibility of overfitting, the number of hidden nodes $s$ need to be small as long as a reasonable level of accuracy can be achieved. In practice, this can be done by monitoring the resulting acceptance rate using an initial chain.

\begin{figure}[!t]
\begin{center}
\includegraphics[width=0.45\textwidth]{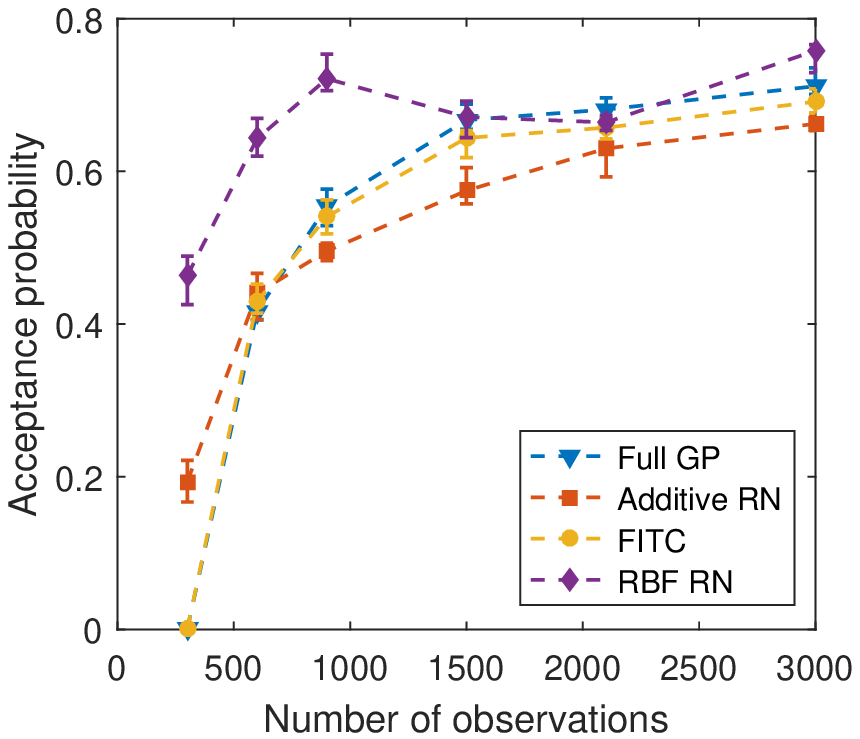} \hspace{-8pt}
\includegraphics[width=0.45\textwidth]{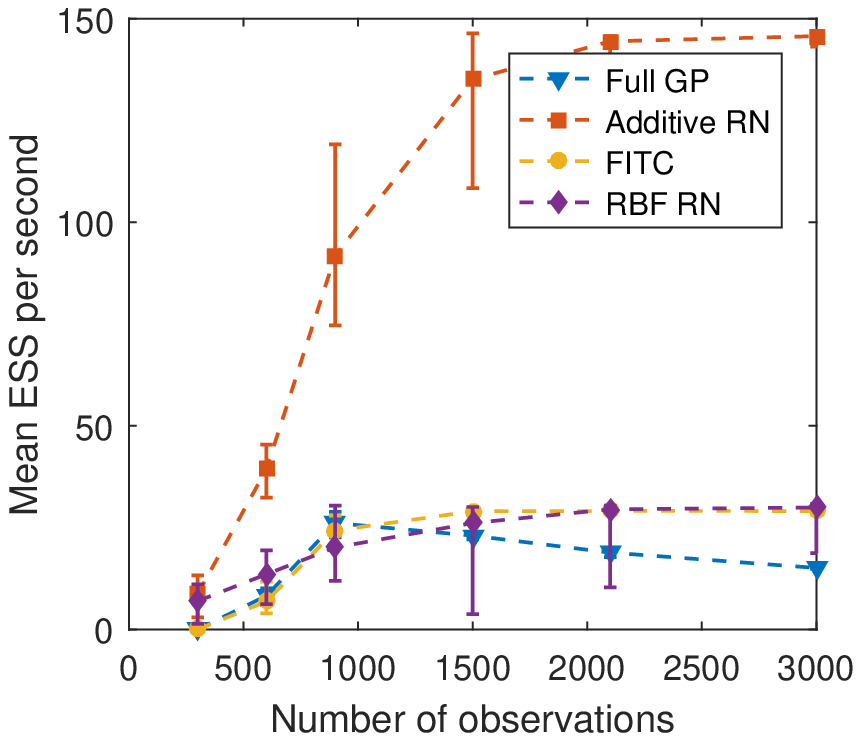}   \hspace{-8pt}
\caption{Comparing the efficiency of our random network surrogates and Gaussian process surrogates on a challenging $32$ dimensional Gaussian target whose covariance matrix has an eigenvector $(1,1,\ldots,1)^T$ with a corresponding eigenvalue of $1.0$, and all other eigenvalues are $0.01$. We set the step size to keep the acceptance probability around $70\%$ for HMC and use the same step size in all surrogate methods.  For FITC and random networks, the number of inducing points and hidden neurons are all set to be $1000$ to allow reasonably accurate approximation. We ran each algorithm ten times and plot the medians and $80\%$ error bars.}\label{fig:gpvsnn}
\vspace{-10pt}
\end{center}
\end{figure} 

\begin{algorithm}[t]
{\small
\KwIn{Starting position $q^{(1)}$, step size $\epsilon$ and number of hidden units $s$}
Initialize the training data set: $D=\emptyset$ or several random samples from the prior\\
 \For{$t =1,2,\cdots, B$ }{
  \textit{Resample momentum $p$}\\
  $p^{(t)} \sim \mathcal{N}(0,M),\;(q_0,p_0)$ = $(q^{(t)},p^{(t)})$\\
  \textit{Simulate discretization of Hamiltonian dynamics and propose $(q^{\ast},p^{\ast})$}\\
  \textit{Metropolis-Hasting correction:} \\
  $u \sim \text{Uniform}[0,1]$,  $\rho = \exp[{ H(q^{(t)},p^{(t)})-H(q^{\ast},p^{\ast}) }]$\\
  \lIf{$u < \min(1,\rho)$,} {$q^{(t+1)} = q^{\ast},\; D = D\cup\{(q^{\ast},U(q^{\ast}))\}$}
 }
 Train a neural network with $s$ hidden units via ELM on $D$ to form the surrogate function $z$\\
 \For{$t =B+1,B+2,\cdots $ }{
  \textit{Resample momentum $p$}\\
  $p^{(t)} \sim \mathcal{N}(0,M),\;(q_0,p_0)$ = $(q^{(t)},p^{(t)})$\\
  \textit{Simulate discretization of a new Hamiltonian dynamics using $z$:}\\
  \For{$l = 1$ to $L$} {
  $p_{l-1} \leftarrow p_{l-1} - \frac{\epsilon}{2} \frac{\partial z}{\partial q}(q_{l-1})$\\
  $q_l \leftarrow q_{l-1} + \epsilon M^{-1}p_{l-1}$\\
   $p_l \leftarrow p_l - \frac{\epsilon}{2} \frac{\partial z}{\partial q}(q_{l})$
  }
  $(q^{\ast},p^{\ast}) = (q_L,p_L)$\\
  \textit{Metropolis-Hasting correction:} \\
  $u \sim \text{Uniform}[0,1]$, $\rho = \exp[{ H(q^{(t)},p^{(t)})-H(q^{\ast},p^{\ast}) }]$\\
  \lIf{$u < \min(1,\rho)$,} {$q^{(t+1)} = q^{\ast}$}
 }

\caption{Random Network Surrogate HMC}
}
\label{alg:RNSHMC}
\end{algorithm}

Following \cite{rasmussen03}, we propose to run our method, henceforth called random network surrogate Hamiltonian Monte Carlo (RNS-HMC, see Algorithm \ref{alg:RNSHMC}), in two phases: exploration phase and exploitation phase. During the exploration phase, we initialize the training data set $D$ with an empty set or some samples from the prior distribution of parameters. We then run the standard HMC algorithm for some iterations and collect information from the new states (i.e., accepted proposals). When we have sufficiently explored the high density domain in parameter space and collected enough training data (during the burn-in period), a shallow random network is trained based on the collected training set $D$ to form a surrogate for the potential energy function. The surrogate function will be used to approximate the gradient information needed for HMC simulations later in the exploitation phase. 

As an illustrative example, we compare the performance of different surrogate HMC methods on a challenging Gaussian target density in $32$ dimensions (A lower dimensional case was used in \cite{rasmussen03}). The target density has 31 confined directions and a main direction that is $10$ times wider, and all variables are correlated. Both the full GPs and FITC methods are implemented using GPML package \cite{rasmussen10}. The results are presented in Figure \ref{fig:gpvsnn}. Compared to the full GPs, FITC and random networks (with additive and RBF nodes) all scales linearly with the number of observations. Both random network surrogates can start with fewer training data. We also compare the efficiency of the surrogate induced Hamiltonian flows in terms of time normalized mean effective sample sizes (ESS). The efficiency of FITC and random networks all increases as the number of observations increase until no more approximation gain can be obtained (see the acceptance probability in the middle panel). However, the efficiency of full GP begin to drop before the model reaches its full capacity. That is because its predictive complexity also grows with the number of observations, which in turn diminishes the overall efficiency. Overall, the random network with additive nodes outperform other methods based on these example. 

Our proposed method provides a natural framework to incorporate surrogate functions in HMC. Moreover, it can be easily extended to RMHMC. To this end, the Hessian matrix of the surrogate function can be used to construct a metric in parameter space and the third order derivatives can be used to simulate the corresponding modified Hamiltonian flow. We refer to this extended version of our method as RNS-RMHMC. 

\begin{figure}[!t]
\begin{center}
\includegraphics[width=0.34\textwidth,height=0.25\textheight]{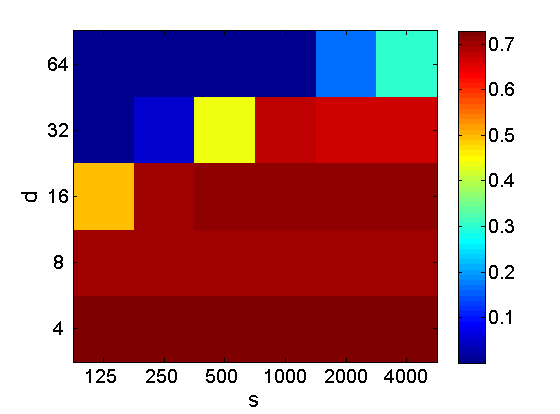} \hspace{-12pt}
\includegraphics[width=0.34\textwidth,height=0.25\textheight]{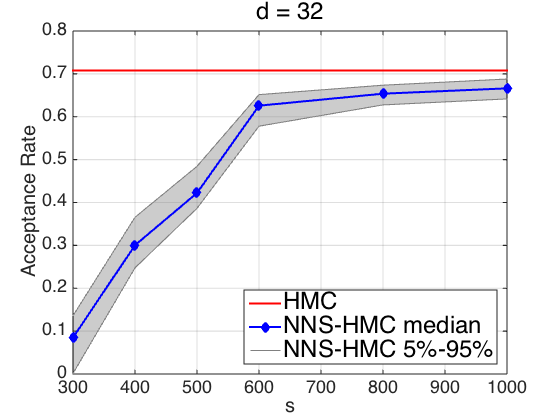} \hspace{-12pt}
\includegraphics[width=0.34\textwidth,height=0.25\textheight]{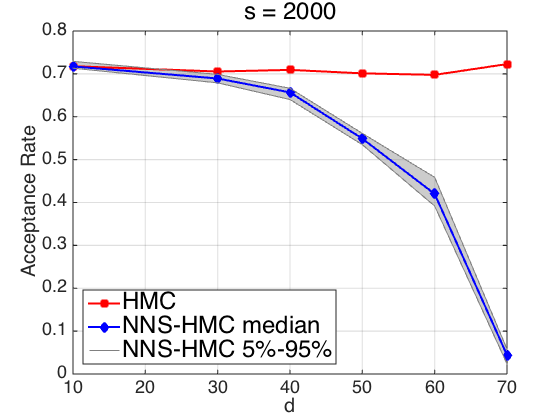} 
\caption{Acceptance probability of surrogate induced Hamiltonian flow on simulated logistic regression models for different number of parameters, $d$, and hidden neurons, $s$. The step size is chosen to keep the acceptance probability around $70\%$ for HMC. {\bf Left:} Acceptance probability as a function of $s$ (x-axis) and $d$ (y-axis). {\bf Middle:} Acceptance probability as a function of $s$ for a fixed dimension $d=32$. {\bf Right:} Acceptance probability as a function of $d$ for a fixed $s=2000$.}\label{fig:dimvsneuron}
\vspace{-10pt}
\end{center}
\end{figure} 

Note that the approximation quality of the neural network surrogate function depends on several factors including the dimension of parameter space, $d$, the number of hidden neurons, $s$ and the training size, $N$. Here, we assume that $N$ is sufficiently large enough, and investigate the efficiency of RNS-HMC in terms of its acceptance probability for different values of $d$ and $s$ based on a standard logistic regression model with simulated data. Similar to the results presented in \cite{strathmann15}, Figure \ref{fig:dimvsneuron} shows the acceptance rate (over $10$ MCMC runs) as a function of $d$ and $s$. For dimensions up to $d\approx 50$, RNS-HMC maintains a relatively high acceptance probability with a shallow random network trained in a few seconds on a laptop computer. Appendix \ref{sec:potentialmatching} provides a theoretical justification for our method.

\section{Adaptive RNS-HMC}\label{sec:OLRNSHMC}
So far, we have assumed that the neural network model in our method is trained using a sufficiently large enough number of training points after waiting for an adequate number of iterations to allow the sampler explore the parameter space. This, however, could be very time consuming in practice: waiting for a long time to collect a large number of training points could undermine the benefits of using the surrogate Hamiltonian function.

\begin{figure}[!t]
\begin{center}
\includegraphics[width=0.34\textwidth,height=0.25\textheight]{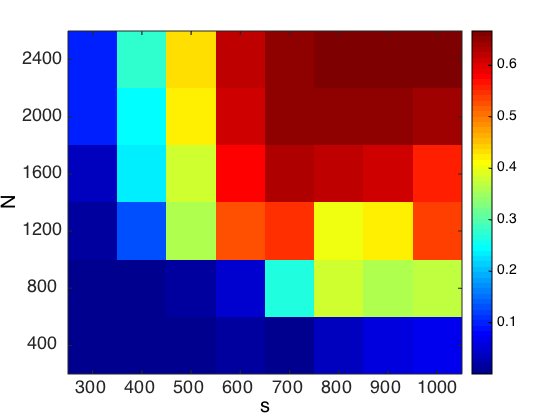} \hspace{-12pt}
\includegraphics[width=0.34\textwidth,height=0.25\textheight]{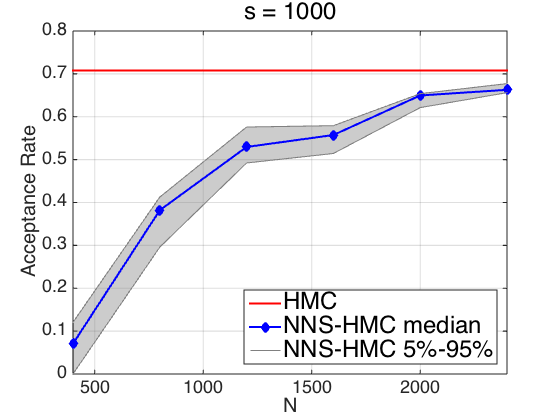}  \hspace{-12pt}
\includegraphics[width=0.34\textwidth,height=0.25\textheight]{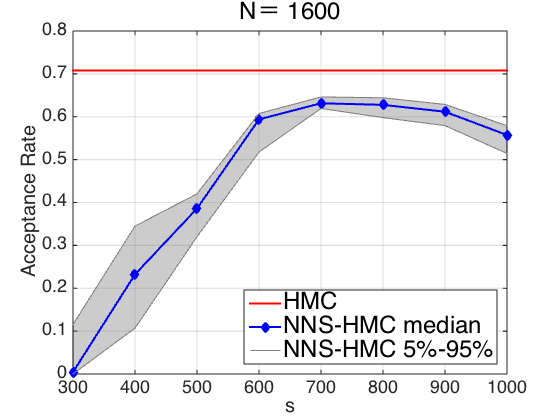}  
\caption{Acceptance probability of the surrogate induced Hamiltonian flow based on a simulated logistic regression models with dimension $d=32$. {\bf Left:} Acceptance probability as a function of the number of hidden neurons $s$ (x-axis) and the number of training points $N$ (y-axis). {\bf Middle:} Acceptance probability as a function of $N$ for a fixed $s=1000$ . {\bf Right:} Acceptance probability as a function of $s$ for a fixed $N=1600$ .}\label{fig:ntrainvsneuron}
\vspace{-10pt}
\end{center}
\end{figure}

Figure \ref{fig:ntrainvsneuron} shows the average acceptance probabilities (over $10$ MCMC chains) as a function of the number of training points, $N$, and the number hidden neurons, $s$, on a simulated logistic regression model for a fixed number of parameters, $d=32$. While it takes around $2000$ training data points to fulfill the network's capability and reach a high acceptance comparable to HMC, only $500$ training points are enough to provide an acceptable surrogate Hamiltonian flow (around $0.1$ acceptance probability). Therefore, we can start training the neural network surrogate earlier and adapting it as more training points become available. Although adapting a Markov chain based on its history may undermine its ergodicity and consequently its convergence to the target distribution \cite{andrieu08}, we can enforce a vanishing adaption rate $a_t$ such that $a_t\rightarrow 0$ and $\sum_{t=1}^\infty a_t = \infty$ and update the surrogate function with probability $a_t$ at iteration $t$. By Theorem 5 of \cite{roberts07}, the resulting algorithm is ergodic and converges to the right target distribution. 

It is straightforward to adapt the neural network surrogate from the history of Markov chain. However, the
estimator in Algorithm \ref{alg:ELM} costs $\mathcal{O}(Nds+Ns^2+s^3)$ computation and $\mathcal{O}(Ns)$ storage, where $N$ is the number of training data (e.g., the number of rows in the output matrix $H$). As $N$ increases, finding the right weight and bias for the output neuron becomes increasingly difficult. In \cite{greville60}, Grevillle shows that $H^{\dagger}$ can be learned incrementally in real time as new data points become available. Based on Greville's method, online and adaptive pseudoinverse solutions for updating ELM weights has been proposed in \cite{vanschaik15} which can be readily employed here to develop an adaptive version of RNS-HMC. To be more efficient, only the estimator is updated.

\begin{proposition}\label{prop:online}
Suppose the current output matrix is $H_k$ and the target vector is $T_k$. At time $k+1$, a new sample $q_{k+1}$ and the target (potential) $t_{k+1}$ are collected. Denote the output vector from the hidden layer at $q_{k+1}$ as $h_{k+1}$. The adaptive updating formula for the empirical potential matching estimator is given by 
\[
v_{k+1} = v_{k} + (t_{k+1}-h_{k+1}^Tv_{k})b_{k+1}
\]
where $b_{k+1}$ and auxiliary matrices $\Phi_{k+1},\;\Theta_{k+1}$ are updated according to $c_{k+1} = \Phi_{k}h_{k+1}$.
\begin{itemize}
\item[(i)] $c_{k+1} = 0$
\[
b_{k+1} =\frac{\Theta_{k}h_{k+1}}{1+ h_{k+1}^T\Theta_{k}h_{k+1}}, \quad \Phi_{k+1} = \Phi_k, \quad \Theta_{k+1} = \Theta_k - \Theta_{k}h_{k+1}b_{k+1}^T
\]
\item[(ii)] $c_{k+1}\neq0$
\[
b_{k+1} =\frac{ c_{k+1}}{\|c_{k+1}\|^2}, \quad \Phi_{k+1} = \Phi_k - \Phi_kh_{k+1}b_{k+1}^T
\]
\[
 \Theta_{k+1} = \Theta_k -\Theta_kh_{k+1}b_{k+1}^T + (1+h_{k+1}^T\Theta_kh_{k+1})b_{k+1}b_{k+1}^T - b_{k+1}h_{k+1}^T\Theta_k^T
\]
\end{itemize}
\end{proposition}

At time $k$, the estimator takes a one-off $\mathcal{O}(kds+ks^2+s^3)$ computation and $\mathcal{O}(s^2)$ storage (only need to store $\Phi_k$ and $\Theta_k$, not $H_k^{\dagger}$). Starting at a previously computed solution $v_{K} = H_{K}^{\dagger}T_{K}$, and two auxiliary matrices $\Phi_K=I-H_K^T(H_K^{\dagger})^T,\Theta_K=H_K^{\dagger}(H_K^{\dagger})^T$, this adaptive updating costs $\mathcal{O}(ds + s^2)$ computation and $\mathcal{O}(s^2)$ storage, independent of the training data size $k$. Further details are provided in Appendix \ref{sec:OLproof}. We refer to this extended version of our method as Adaptive RNS-HMC (ARNS-HMC).

\begin{algorithm}[t]
{\small

\KwIn{Initial estimator $v_0$ and auxiliary matrices $\Phi_0,\Theta_0$, adaption schedule $a_t$, }
\myinput{step size $\epsilon$, number of hidden units $s$}

Initialize the surrogate $z_0 = z(q,v_0)$

 \For{$t =0,1,\cdots $ }{
  \textit{Resample momentum $p$}
  
  $p^{(t)} \sim \mathcal{N}(0,M),\;(q_0,p_0)$ = $(q^{(t)},p^{(t)})$
  
  \textit{Propose $(q^{\ast},p^{\ast})$ by simulating discretization of Hamiltonian dynamics with $\frac{\partial U}{\partial q} = \frac{\partial z_t}{\partial q}$}

  \textit{Metropolis-Hasting correction:}

  $u \sim \text{Uniform}[0,1]$,  $\rho = \exp[{ H(q^{(t)},p^{(t)})-H(q^{\ast},p^{\ast}) }]$

  \lIf{$u < \min(1,\rho)$,} {$q^{(t+1)} = q^{\ast}$;} \lElse{$q^{(t+1)} = q^{(t)}$}

  update the estimator and auxiliary matrices to $v_{t+1},\Phi_{t+1},\Theta_{t+1}$ using ($q^{(t+1)},U(q^{(t+1)})$)

  $u \sim \text{Uniform}[0,1]$, 
  
  \lIf{$u < a_t$,} {$z_{t+1} = z(q,v_{t+1})$;} \lElse{$z_{t+1} = z_t$}
 }
 \caption{ARNS-HMC}
}
\label{alg:OLNNSHMC}
\end{algorithm}

\section{Experiments}
\label{sec:results}
In this section, we use several experiments based on logistic regression models and inverse problem for elliptic partial differential equation (PDE) to compare our proposed methods to standard HMC and RMHMC in terms of sampling efficiency defined as time-normalized effective sample size (ESS). Given $B$ MCMC samples for each parameter, ESS =
$B[1 + 2\Sigma_{k=1}^{K}\gamma(k)]^{-1}$, where $\Sigma_{k=1}^{K}\gamma(k)$ is
the sum of $K$ monotone sample autocorrelations \cite{geyer92}. We use the
minimum ESS over all parameters normalized by the CPU time, $s$ (in seconds), as the overall measure
of efficiency: $\min(\textrm{ESS})/\textrm{s}$. The corresponding step sizes $\epsilon$ and number of leapfrog steps $L$ for HMC and RMHMC are chosen to make them stable and efficient (e.g., reasonably high acceptance probability and fast mixing).  The same settings are used for our methods. Note that while the acceptance rates are similar in the first two examples, they drop a little bit for the last two examples, which is mainly due to the constraints we imposed on our surrogate functions. To prevent non-ergodicity and ensure high ESS for both HMC and RNS-HMC, we follow the suggestion by \cite{neal11} to uniformly sample $L$  from $\{1, \ldots,L\}$ in each iteration. The number of hidden nodes $s$ in random network surrogates are not tuned too much and better results could be obtained by more careful tunings.

In what follows, we first compare different methods in terms of their time-normalized ESS after the burin-period. To this end, we collect $5000$ samples after a reasonable large number of iterations ($5000$) of burn-in to make sure the chains have reached their stationary states. For our methods, the accepted proposals during the burn-in period after a short warming-up session (the first 1000 iterations) are used as a training set for a shallow random network. Later, we show the advantages of our adaptive algorithm.


\subsection{Logistic regression model}
As our first example, we compare HMC, RMHMC, RNS-HMC, and RNS-RMHMC using a simulation study based on a logistic regression model with $50$ parameters and $N=10^5$ observations. The design matrix is $X=(\frac1{10}\vect{1},X_1)$ and true parameters $\beta$ are uniformly sampled from $[0,1]^{50}$, where $X_1\sim\mathcal{N}_{49}(\vect{0},\frac1{100}I_{49})$. The binary responses $Y=(y_1,y_2,\ldots,y_N)^T$ are sampled independently from Bernoulli distributions with probabilities $p_i = 1/(1+\exp(-x_i^T\beta))$. We assume $\beta\sim\mathcal{N}_{50}(\vect{0},100I_{50})$, and sample from the corresponding posterior distribution. 

Notice that the potential energy function $U$ is now a convex function, the Hessian matrix is positive semi-definite everywhere. Therefore, we use the Hessian matrix of the surrogate as a local metric in RNS-RMHMC. For HMC and RNS-HMC, we set the step size and leapfrog steps $\epsilon = 0.045,\;L=6$. For RMHMC and RNS-RMHMC, we set the step size and leapfrog steps $\epsilon = 0.54,\;L=2$.

\begin{figure}[!t]
\begin{tabular}{cc}
\includegraphics[width=0.45\textwidth]{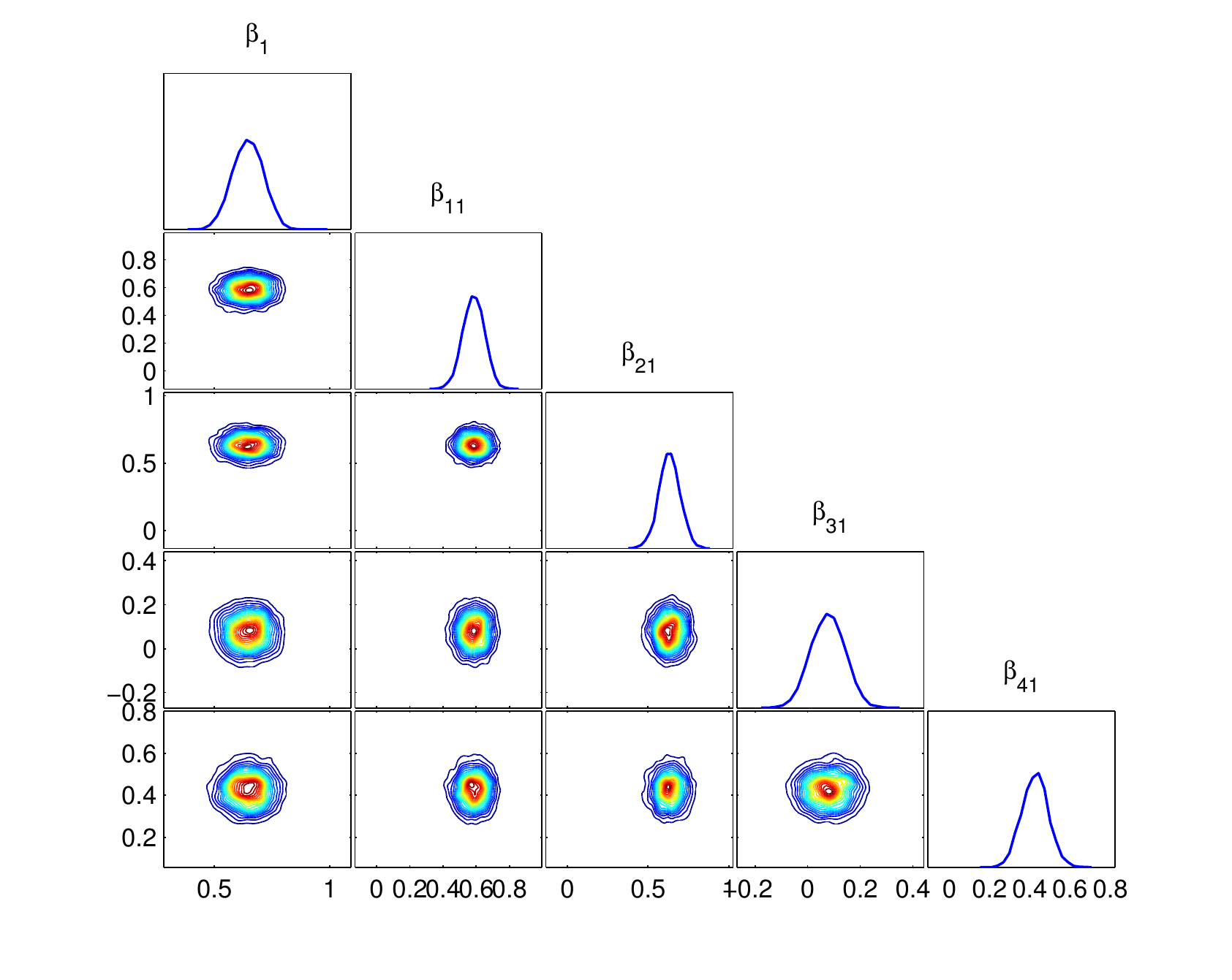}&
\includegraphics[width=0.45\textwidth]{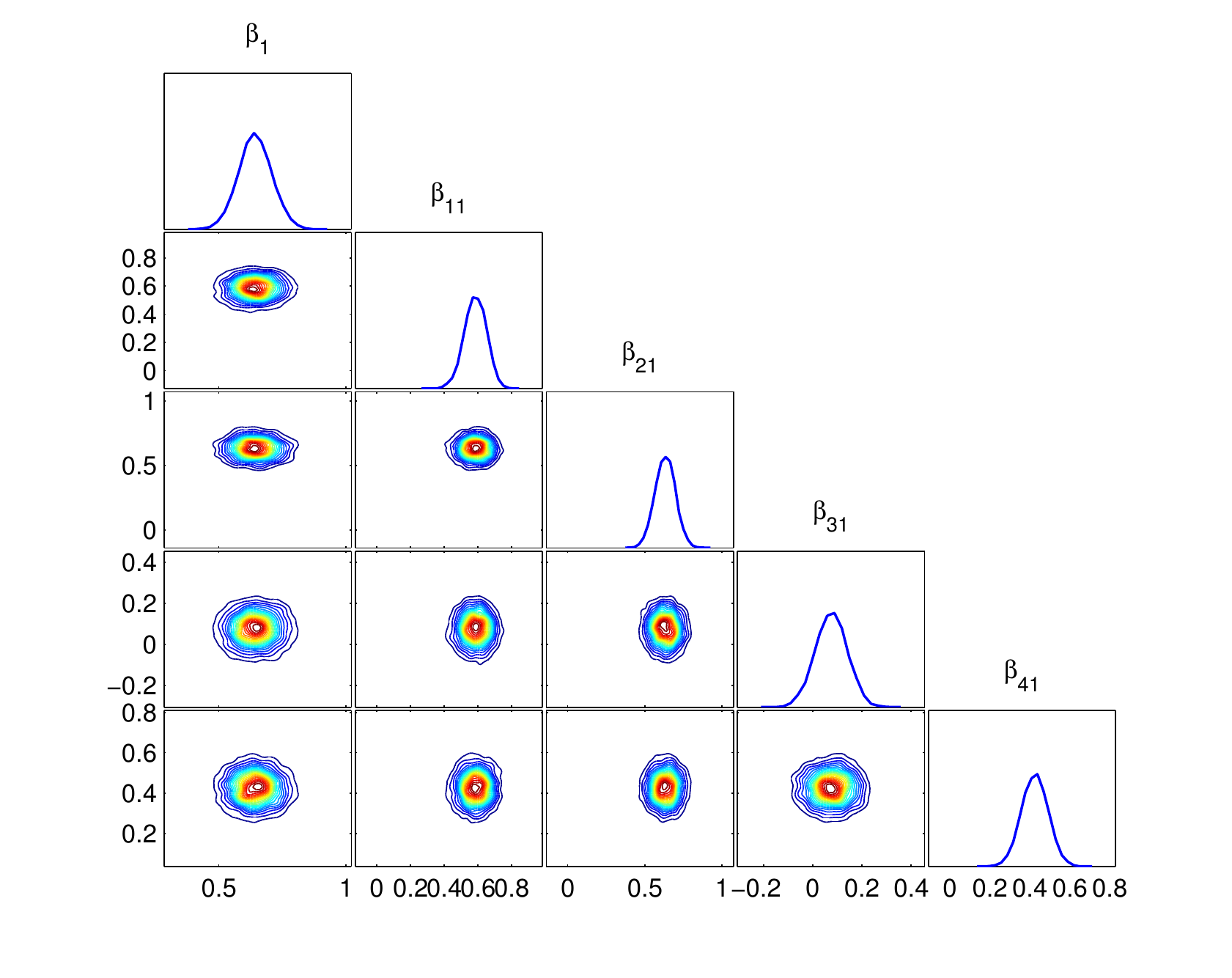} \\
HMC  &  RNS-HMC  
\end{tabular}
\caption{HMC vs RNS-HMC: Comparing one- and two-dimensional posterior marginals of $\beta_1,\beta_{11},\beta_{21},\beta_{31},\beta_{41}$ based on the logistic regression model with simulated data.}\label{fig:logit}
\end{figure}

To illustrate that our method indeed converges to the right target distribution, Figure \ref{fig:logit} provides the one- and two-dimensional posterior marginals of some selected parameters obtained by HMC and RNS-HMC. Table \ref{tab:experiments} compares the performance of the four algorithms. As we can see, RNS-HMC has substantially improved the sampling efficiency in terms of time-normalized min(ESS). 

\begin{table}\begin{center}
{\small
\begin{tabular}{|c|c|c|c|c|c|c|}\hline
Experiment & Method & AP & ESS&  s/Iter & $\min(\textrm{ESS})/\textrm{s}$ & spdup \\\hline
\multirow{2}{*}{LR (Simulation)} & HMC    
& 0.76  & (4351,5000,5000) & 0.061 & 14.17 & 1  \\
& RMHMC & 0.80      &  (1182,1496,1655) & 3.794  & 0.06  & 0.004 \\
\multirow{2}{*}{$s$ = 2000}& RNS-HMC & 0.76   & (4449,4999,5000) & 0.007 & 123.56 & \bf{8.72} \\
& RNS-RMHMC & 0.82  & (1116,1471,1662)  & 0.103 & 2.17 & 0.15 \\\hline
\multirow{2}{*}{LR (Bank Marketing)} & HMC     
& 0.70  & (2005,2454,3368) & 0.061 & 6.52  & 1 \\
& RMHMC & 0.92 & (1769,2128,2428) & 0.631 & 0.56 & 0.09 \\
\multirow{2}{*}{$s$ = 1000}& RNS-HMC & 0.70 & (1761,2358,3378)  & 0.007 & 52.22 & \bf{8.01} \\
& RNS-RMHMC & 0.90  & (1974,2254,2457) & 0.027 & 14.41 & 2.21 \\\hline
\multirow{2}{*}{LR ($\mathtt{a9a}$ 60 dimension)} & HMC     
& 0.72  & (1996,2959,3564)  & 0.033 & 11.96  & 1 \\
& RMHMC & 0.82 & (5000,5000,5000) & 3.492 & 0.29 & 0.02 \\
\multirow{2}{*}{$s$ = 2500}& RNS-HMC & 0.68 & (1835,2650,3203) & 0.005 & 81.80 & \bf{6.84} \\
& RNS-RMHMC & 0.79 & (4957,5000,5000) & 0.370 & 2.68 & 0.22 \\\hline
\multirow{2}{*}{Elliptic PDE} &  HMC     
& 0.91  & (4533,5000,5000)  & 0.775 & 1.17  & 1 \\
& RMHMC & 0.80 & (5000,5000,5000) & 4.388 & 0.23 & 0.20 \\
\multirow{2}{*}{$s$ = 1000}
&RNS-HMC  & 0.75  & (2306,3034,3516) & 0.066 & 7.10 & \bf{6.07} \\
& RNS-RMHMC & 0.66 & (2126,4052,5000) & 0.097 & 4.38 & 3.74 \\\hline
\end{tabular}
\caption{\small Comparing the algorithms using logistic regression models and an elliptic PDE inverse problem. For each method, we provide the acceptance probability (AP), the CPU time (s) for each iteration and the time-normalized ESS.}\label{tab:experiments}

\vspace{-10pt}
}
\end{center}
\end{table}

Next, we apply our method to two real datasets: Bank Marketing and the $\mathtt{a9a}$ dataset \cite{lin08}. The Bank Marketing dataset (40197 observations and 24 features) is collected based on direct marketing campaigns of a Portuguese banking institution aiming at predicting if a client will subscribe to a term deposit \cite{moro14}. We set the step size and number of leapfrog steps $\epsilon=0.012,\;L=45$ for HMC and RNS-HMC; $\epsilon=0.4,\;L=6$ for RMHMC and RNS-RMHMC. The $\mathtt{a9a}$ dataset (32561 features and 123 features) is complied from the UCI $\mathtt{adult}$ dataset \cite{bache13} which has been used to determine whether a person makes over 50K a year. We reduce the number of features to $60$ by random projection (increasing the dimension to 100 results in a substantial drop in the acceptance probability). We set the step size and number of leapfrog steps $\epsilon=0.012,\;L=10$ for HMC and RNS-HMC; $\epsilon=0.5,\;L=4$ for RMHMC and RNS-RMHMC. All datasets are normalized to have zero mean and unit standard deviation. The priors are the same as before. The results for the two data sets are summarized in Table 1. As before, both RNS-HMC and RNS-RMHMC significantly outperform their counterpart algorithms.

\subsection{Elliptic PDE inverse problem}
\label{sec:PDE}
Another computationally intensive model is the elliptic PDE inverse problem discussed in \cite{dashti11,conard14}. This classical inverse problem involves inference of the diffusion coefficient in an elliptic PDE which is usually used to model isothermal steady flow in porous media. Let $c$ be the unknown diffusion coefficient and $u$ be the pressure field, the forward model is governed by the elliptic PDE
\begin{equation}\label{eq:ePDE}
\nabla_{\vect{x}}\cdot(c(\vect{x},\theta)\nabla_{\vect{x}}u(\vect{x},\theta)) = 0,
\end{equation}
where $\vect{x} =(x_1,x_2)\in[0,1]^2$ is the spatial coordinate. The boundary conditions are
\[
u(\vect{x},\theta)|_{x_2=0} = x_1,\quad u(\vect{x},\theta)|_{x_2=1} = 1-x_1,
\quad
\frac{\partial u(\vect{x},\theta)}{\partial x_1}\Big|_{x_1=0} = 0,\quad \frac{\partial u(\vect{x},\theta)}{\partial x_1}\Big|_{x_1=1} = 0
\]

In our numerical simulation, \eqref{eq:ePDE} is solved using standard continuous GFEM with bilinear basis functions on a uniform $30\times30$ quadrilateral mesh. 

A log-Gaussian process prior is used for $c(\vect{x})$ with mean zero and an isotropic squared-exponential covariance kernel:
\[
C(\vect{x}_1,\vect{x}_2) = \sigma^2\exp\left(-\frac{\|\vect{x}_1-\vect{x}_2\|^2_2}{2l^2}\right)
\]
for which we set the variance $\sigma^2 =1$ and the length scale $l=0.2$. Now, the diffusivity field can be easily parameterized with a Karhunen-Loeve (K-L) expansion:
\[
c(\vect{x},\theta) \approx \exp\left(\sum_{i=1}^d\theta_i\sqrt{\lambda_i}v_i(\vect{x})\right)
\]
where $\lambda_i$ and $v_i(\vect{x})$ are the eigenvalues and eigenfunctions of the integral operator defined by the kernel $C$, and the parameter $\theta_i$ are endowed with independent standard normal priors, $\theta_i\sim\mathcal{N}(0,0.5^2)$, which are the targets of inference. In particular, we truncate the K-L expansion at $d=20$ modes and condition the corresponding mode weights on data. Data are generated by adding independent Gaussian noise to observations of the solution field on a uniform $11\times 11$ grid covering the unit square.
\[
y_j = u(\vect{x}_j,\theta) + \epsilon_j,\quad \epsilon_j\sim\mathcal{N}(0,0.1^2), \quad j = 1,2,\ldots,N
\]

The number of leap frog steps and step sizes are set to be $L=10,\; \epsilon = 0.16$ for both HMC and NNS-HMC. Note that the potential energy function is no longer convex; therefore, we can not construct a local metric from the Hessian matrix directly. However, the diagonal elements 
\[
\frac{\partial^2 U}{\partial \theta_i^2} = \frac{1}{\sigma_\theta^2} + \sum_{j=1}^{N}\frac{1}{\sigma_y^2}\left(\frac{\partial u_j}{\partial \theta_i}\right)^2 - \sum_{j=1}^{N}\frac{\epsilon_j}{\sigma_y^2}\frac{\partial^2 u_j}{\partial \theta_i^2}, \quad \sigma_\theta = 0.5,\;\sigma_y = 0.1,\quad i=1,2,\ldots,d
\]
are highly likely to be positive in that the deterministic part (first two terms) is always positive and the noise part (last term) tends to cancel out. The diagonals of the Hessian matrix of surrogate therefore induce an effective local metric which can be used in RNS-RMHMC. A comparison of the results of all algorithms are presented in Table \ref{tab:experiments}. As before, RNS-HMC provides a substantial improvement in the sampling efficiency. For the RMHMC methods, we set $L=3,\;\epsilon=0.8$. As seen in the table, RMHMC is less efficient than HMC mainly due to the slow computation speed. However, RNS-RMHMC improves RMHMC substantially and  outperforms HMC. Although the metric induced by the diagonals of the Hessian matrix of surrogate may not be as effective as Fisher information, it is much cheaper to compute and provide a good approximation. 

{\bf Remark.} In addition to the usual computational bottleneck as in previous examples, e.g., large amount of data, there is another challenge on top of that for this example due to the complicated forward model. Instead of a simple explicit probabilistic model that prescribes the likelihood of data given the parameter of interest, a PDE \eqref{eq:ePDE} is involved in the probabilistic model. The evaluation of geometrical and statistical quantities, therefore, involves solving a PDE similar to \eqref{eq:ePDE} in each iteration of HMC and RHMHC. This is a preventive factor in practice. Using our methods based on neural network surrogates provide a huge advantage. Numerical experiments show a gain of efficiency by more than 20 times. More improvement is expected as the amount of data increases. 

\begin{figure}[!t]
\begin{center}
\begin{tabular}{cc}
\includegraphics[width=0.45\textwidth,height=0.25\textheight]{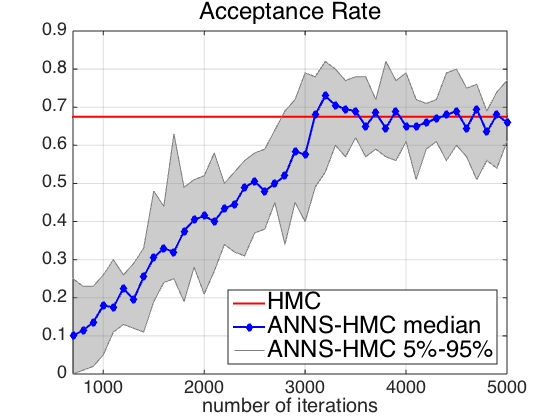}  & 
\includegraphics[width=0.45\textwidth,height=0.25\textheight]{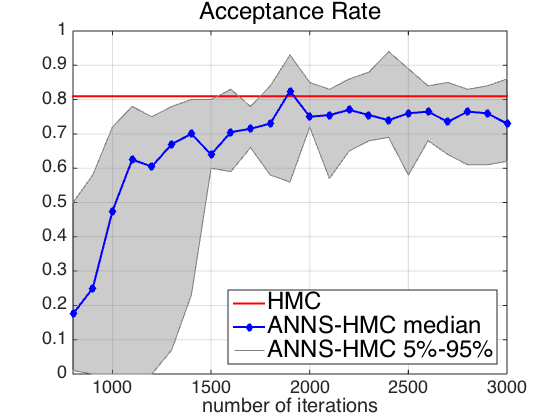} \\
(a) Simulated Data & (b) Bank Market\\ \\
\includegraphics[width=0.45\textwidth,height=0.25\textheight]{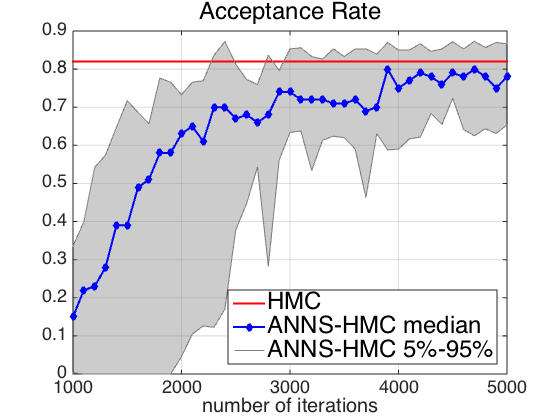}  & 
\includegraphics[width=0.45\textwidth,height=0.25\textheight]{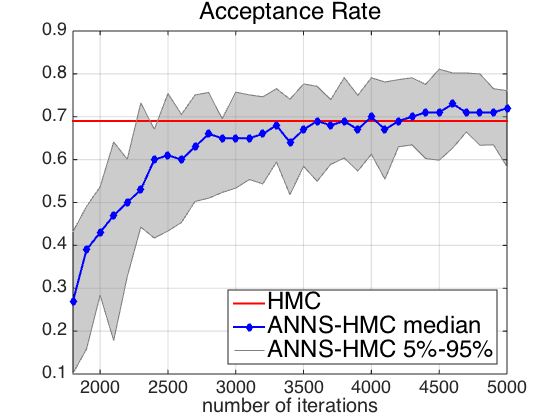}  \\
(c) Adult Data & (d) Elliptic PDE
\end{tabular}
\caption{Median acceptance rate of ANNS-HMC along with the corresponding 90\% interval (shaded area). The red line shows the average acceptance rate of standard HMC.} \label{fig:acprat}
\end{center}
\end{figure}
\begin{figure}[!t]
\begin{center}
\begin{tabular}{cc}
\includegraphics[width=0.45\textwidth,height=0.25\textheight]{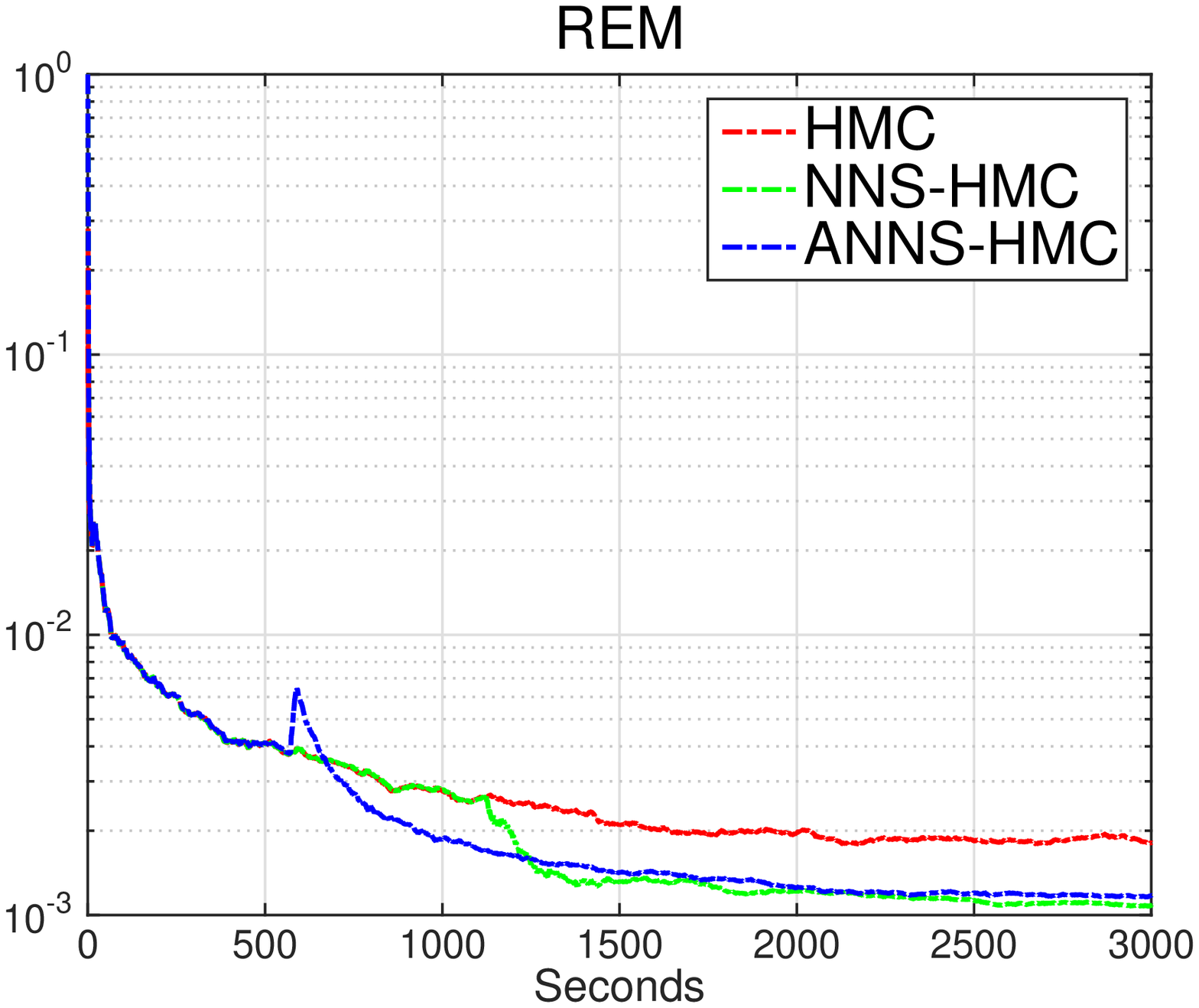}  & 
\includegraphics[width=0.45\textwidth,height=0.25\textheight]{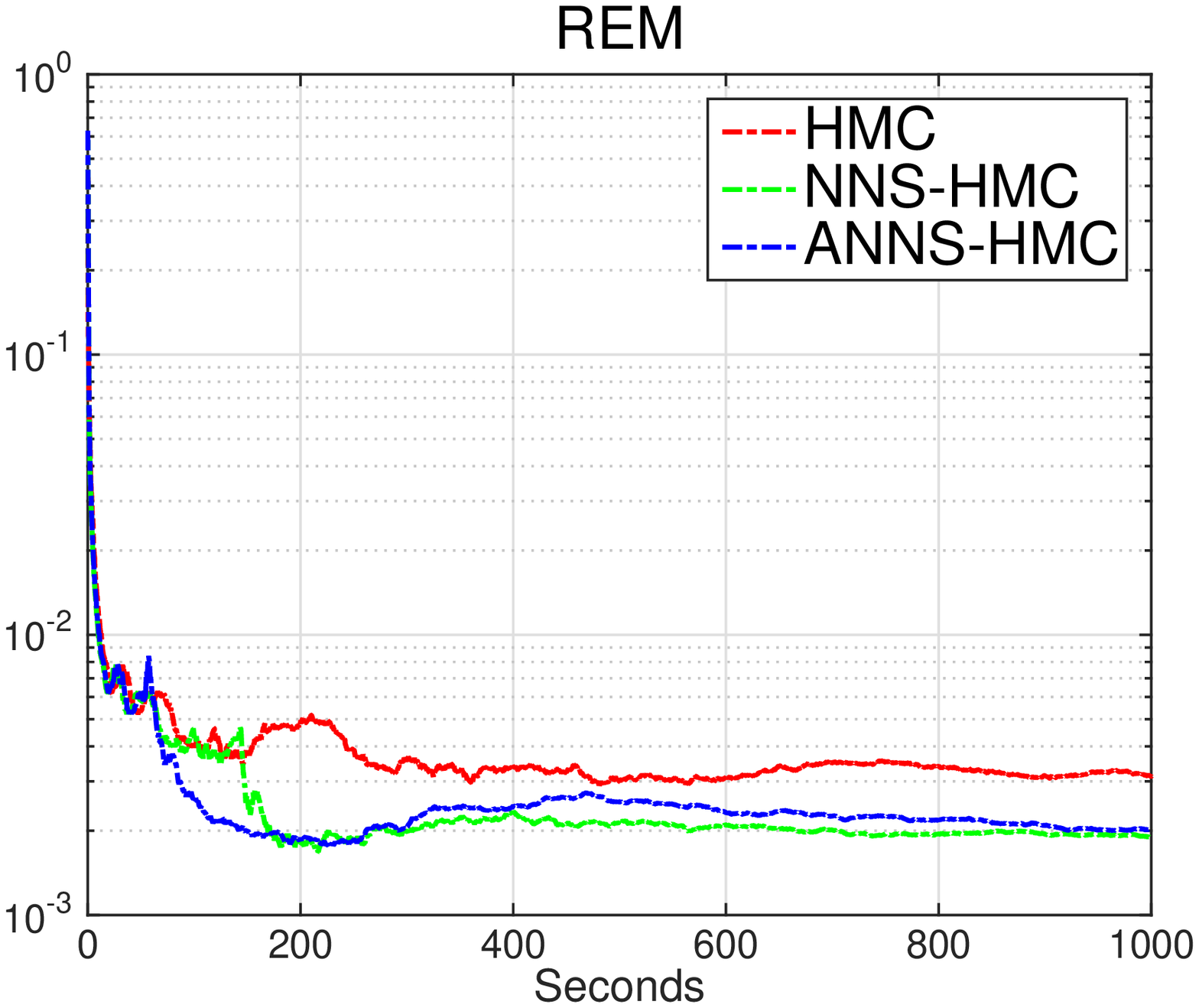} \\
(a) Simulated Data & (b) Bank Market\\ \\
\includegraphics[width=0.45\textwidth,height=0.25\textheight]{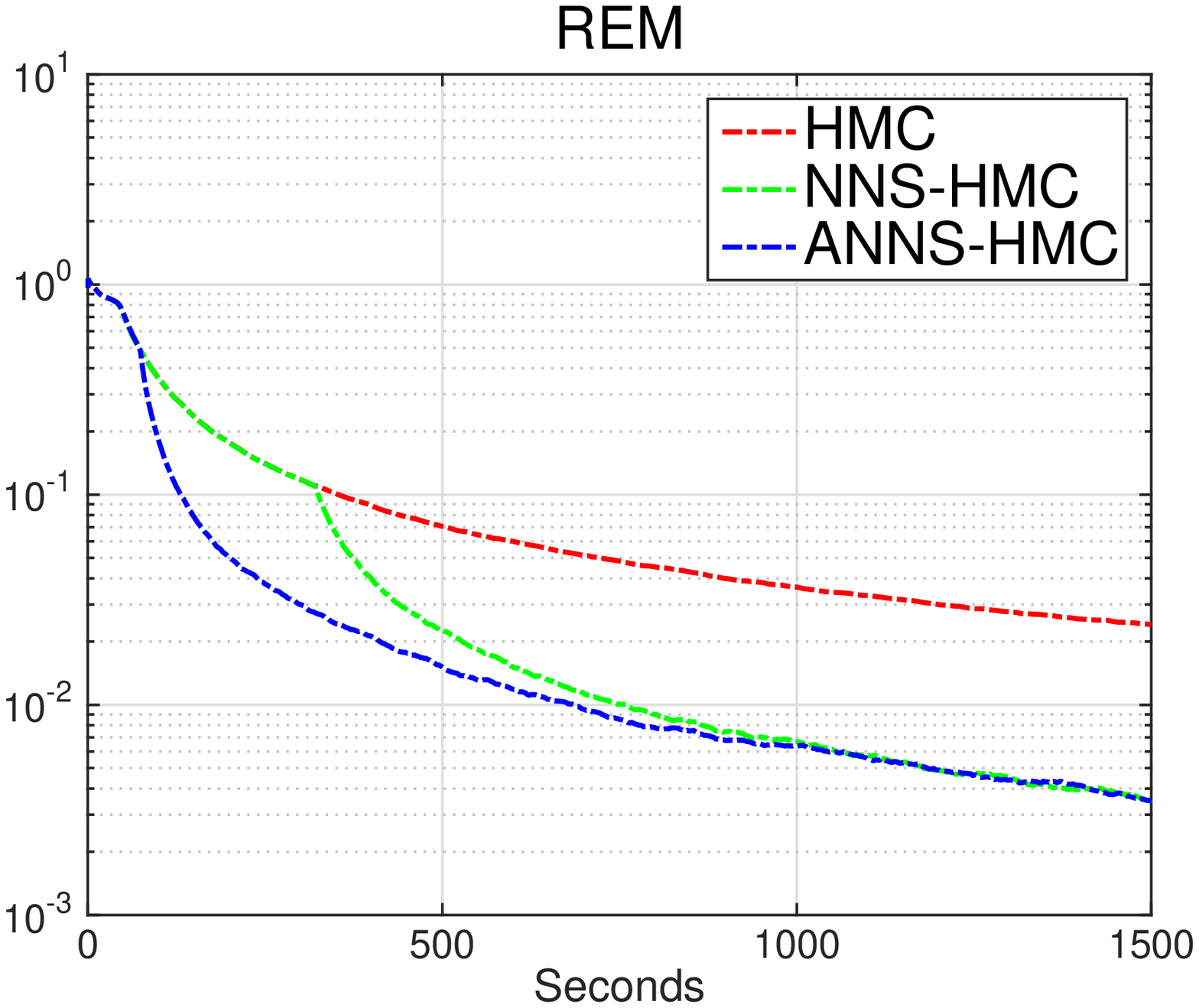}  &
\includegraphics[width=0.45\textwidth,height=0.25\textheight]{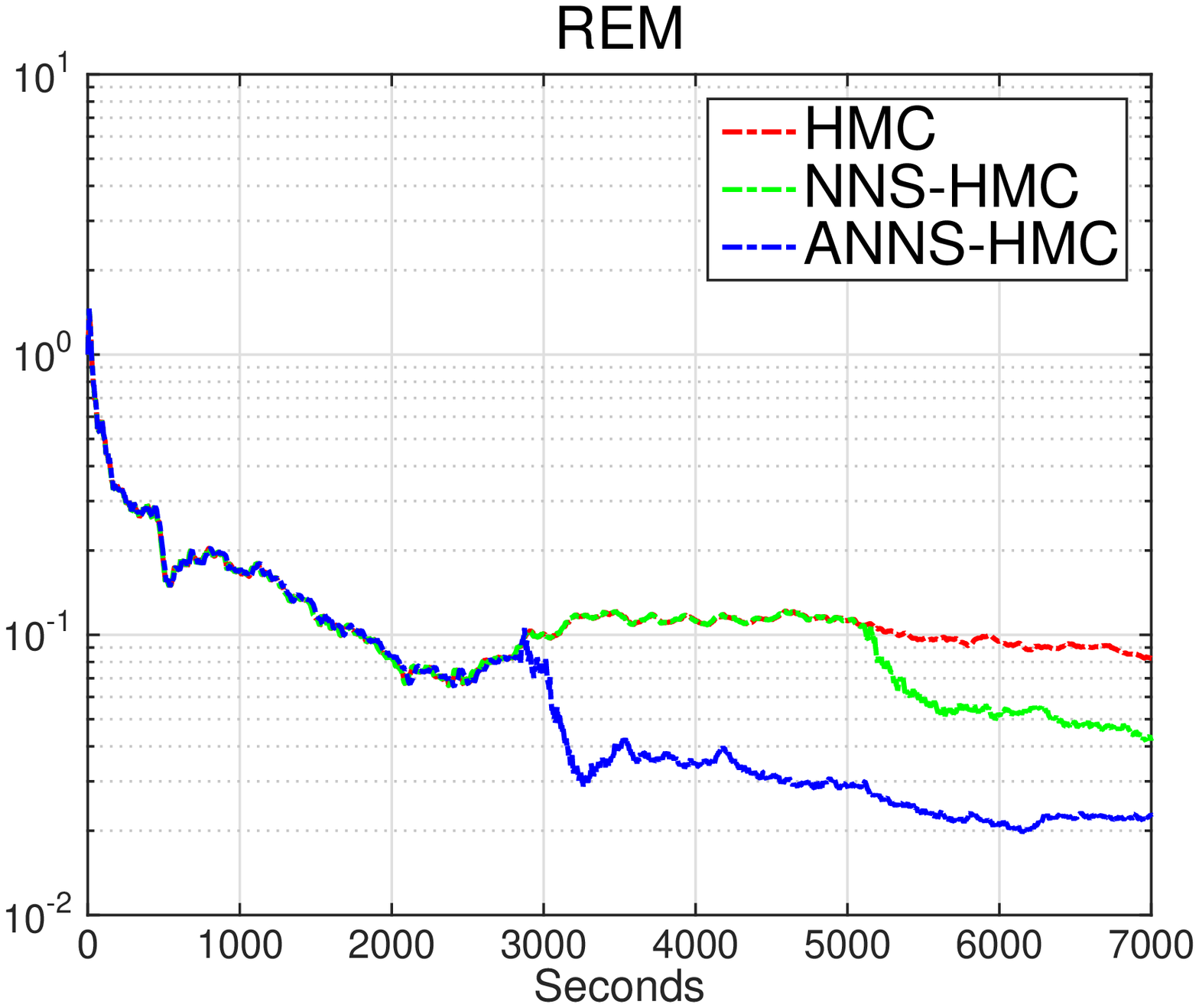}  \\
(c) Adult Data & (d) Elliptic PDE
\end{tabular}
\caption{Relative error of mean as a function of running time.}\label{fig:rem}
\vspace{-10pt}
\end{center}
\end{figure}

\subsection{Adaptive learning}

Next, using the above four examples we show that ARNS-HMC can start with far fewer training points and quickly reach the same level of performance as that of RNS-HMC. Figure \ref{fig:acprat} shows that as the number of training points (from initial MCMC iterations) increases, ARNS-HMC fully achieves the network's capability and reaches a comparable acceptance rate to that of HMC.

We also compare ARNS-HMC to HMC and RNS-HMC in terms of the relative error of mean (REM) which is defined as $\|\bar{q(t)}-E(q)\|_2/\|E(q)\|_2$, where $\bar{q(t)}$ means sample mean up to time $t$. Figure \ref{fig:rem} shows the results using the four examples discussed above. Note that before training the neural network models, both RNS-HMC and ARNS-HMC are simply standard HMC so the three algorithms have similar performance. As we can see, ARNS-HMC has the best overall performance: it tends to provide lower REM at early iterations. This could be useful if we have limited time budget to fit a model.

\section{Discussion and Future Work}
\label{sec:discussion}
In this paper, we have proposed an efficient and scalable computational method for Bayesian inference by exploring and exploiting regularity of probability models in parameter space. Our method is based on training surrogate function of the potential energy after exploring the parameter space sufficiently well. For situations where it is not practical to wait for a thorough exploration of parameter space, we have proposed an adaptive version of our method that can start with fewer training points and can quickly reach its full potential.

As an example, we used random networks and efficient learning algorithms to construct effective surrogate functions. These random bases surrogate functions provide good approximations of collective information of the full data set while striking a good balance between accuracy and computation cost for efficient computation. Random networks combined with the optimized learning process can provide flexibility, accuracy, and scalability. Note that in general the overall performance could be sensitive to the architecture of the random network. Our proposed random network surrogate method scales differently than GP emulators because of the specific constraints we imposed on its architecture. As our experimental results show, this approach could improve the performance of HMC in some applications. 

In its current form, our method is more effective in problems with costly likelihood and a moderate number of parameters. In spite of improvements we have made to standard HMC, dealing with high dimensional and complex distributions still remains quite challenging. For multimodal distributions, for example, our method's effectiveness largely depends on the quality of training samples. If these samples are collected from one mode only, the surrogate function will miss the remaining modes and the sampler might not be able to explore them (especially if they are isolated modes). A surrogate function based on Gaussian processes might have a better chance at finding these modes in the tails of the approximate distribution since it tends to go to zero gradually. To address this issue, we can utilize mode searching and mode exploring ideas such as those proposed by \cite{ahn13, lan14}. For constrained target distributions, we can employ the method of \cite{lanICML14} based on Spherical HMC. 


For HMC, gradient of the potential function is an important driving force in the Hamiltonian dynamics. Although accurate approximation of a well sampled smooth function automatically leads to accurate approximation of its gradient, this is not the case when the sampling is not well distributed. For example, when dense and well sampled training data sets are difficult to obtain in very high dimensions, one can incorporate the gradient information in the training process. In future, we will study more effective way to utilize this information in the training process. As a common practice in adaptive MCMC methods, one may also learn the mass matrix $M$ adaptively together with the surrogate in ARNS-HMC.



\subsubsection*{Acknowledgments}

This work is supported by NIH grant R01AI107034 and NSF grants DMS-1418422 and DMS-1622490



\bibliographystyle{unsrt}
\small{

}

%
%
\newpage
\appendix

\section{Convergence to the correct target distribution}\label{sec:target}
In order to prove that the equilibrium distribution remains the same, it suffices to show that the detailed balance condition still holds. Denote $\theta = (p,q),\; \theta'=\tilde{\phi}_t(\theta)$. In the Metropolis-Hasting step, we use the original Hamiltonian to compute the acceptance probability
\[
\alpha(\theta,\theta') = \min(1,\exp[-H(\theta') + H(\theta)])
\]
therefore,
\begin{align*}
\alpha(\theta,\theta')\mathbb{P}(d\theta) &= \alpha(\theta,\theta') \exp[-H(\theta)]d\theta  \\
&\stackrel{\theta = \tilde{\phi}^{-1}_{t}(\theta')}{=}  \min(\exp[-H(\theta)],\exp[-H(\theta')])\left|\frac{d\theta}{d\theta'}\right|d\theta' \\
& = \alpha(\theta',\theta)\exp[-H(\theta')]d\theta'\\
& = \alpha(\theta',\theta) \mathbb{P}(d\theta')
\end{align*}
since $ \left|\frac{d\theta}{d\theta'}\right| = 1$ due to the volume conservation property of the surrogate induced Hamiltonian flow $\tilde{\phi}_t$. Now that we showed the detailed balance condition is satisfied, along with the reversibility of the surrogate induced Hamiltonian flow, the modified Markov chain will converge to the correct target distribution.

\section{Potential Matching}\label{sec:potentialmatching}
In the paper, training data collected from the history of Markov chain are used without a detailed explanation. Here, we will analyze the asymptotical behavior of surrogate induced distribution and try to explain why the history of the Markov chain is a reasonable choice for training data. Recall that we find our neural network surrogate function by minimizing the mean square error \eqref{eq:cost}. Similarly to \cite{hyvarinen05}, it turns out that minimizing \eqref{eq:cost} is asymptotically equivalent to minimizing a new distance between the surrogate induced distribution and the underlying target distribution, independent of their corresponding normalizing constants.

Suppose we know the density of the underlying intractable target distribution up to a constant 
\[
P(q|Y) = \frac{1}{Z}\exp\left(-U(q)\right)
\]
where $Z$ is the unknown normalizing constant. Our goal is to approximate this distribution using a parametrized density model, also known up to a constant, 
\[
Q(q,\theta) = \frac{1}{Z(\theta)}\exp\left(-V(q,\theta)\right)
\]
Ignoring the multiplicative constant, the corresponding potential energy functions are $U(q)$ and $V(q,\theta)$ respectively. The straightforward square distance between the two potentials will not be a well-defined measure between the two distributions distributions because of the unknown normalizing constants. Therefore, we use the following distance instead:
\begin{align} \label{eq:pm}
K(\theta) &= \min_d\int \|V(q,\theta)-U(q)-d\|^2\;P(q|Y)\;dq\nonumber\\
&=\int \|V(q,\theta)-U(q)\|^2\; P(q|Y)\;dq - [E_q(V(\theta)-U)]^2 = \mathbb{V}\mathrm{ar}_{q}(V(\theta)-U)
\end{align}
Because of its similarity to {\it score matching} \cite{hyvarinen05}, we refer to the approximation method based on this new distance as {\it potential matching}; the corresponding {\it potential matching} estimator of $\theta$ is given by 
\[
\hat{\theta} = \arg\min_{\theta} K(\theta)
\]

It is easy to verify that $K(\theta) = 0 \Rightarrow V(\theta) = U + constant \Rightarrow Q(q,\theta) = P(q|Y)$, so $K(\theta)$ is a well-defined squared distance. Exact evaluation of \eqref{eq:pm} is analytically intractable. In practice, given $N$ samples from the target distribution $q_1,q_2,\ldots,q_N$, we minimize the empirical version of \eqref{eq:pm}
\begin{align}
\tilde{K}(\theta) &= \min_d\frac1N\sum_{n=1}^N \|V(q_n,\theta) - U(q_n)-d\|^2 \nonumber\\
&= \frac1N\sum_{n=1}^N \|V(q_n,\theta)-U(q_n)\|^2 - \left(\frac1N\sum_{n=1}^NV(q_n,\theta)-U(q_n)\right)^2\label{eq:empm}
\end{align}
which is asymptotically equivalent to $K$ by law of large numbers. \eqref{eq:empm} could be more concise if we allow a shift term in the parametrized model ($V(q,\theta) = V(q,\theta_{-d}) + \theta_d$). In that case, the empirical {\it potential matching} estimator is 
\begin{align*}
\tilde{\theta} &= \arg\min_\theta \tilde{K}(\theta) = \arg\min_\theta\min_d\frac1N\sum_{n=1}^N \|V(q_n,\theta_{-d}) +(\theta_d-d) - U(q_n)\|^2\\
&= \arg\min_\theta\frac1N\sum_{n=1}^N \|V(q_n,\theta_{-d}) + \theta_d - U(q_n)\|^2\\
&= \arg\min_\theta\frac1N\sum_{n=1}^N \|V(q_n,\theta) - U(q_n)\|^2
\end{align*}

In particular, if we adopt an additive model for $V(q,\theta)$ with a shift term 
\[
V(q,\theta) = \sum_{i=1}^sv_i\sigma(\vect{w_i}\cdot q+d_i) + b,\quad \theta = (\vect{v},b)
\]
where $\vect{w_i}, d_i$ and activation function $\sigma$ are chosen according to Algorithm \ref{alg:ELM} and collect early evaluations from the history of Markov chain
\[
 \mathcal{T}_N = \{(q^{(1)},U(q^{(1)}),(q^{(2)},U(q^{(2)})),\ldots,(q^{(N)},U(q^{(N)}))\}
 \]
 as training data; this way, the estimated parameters (i.e., weights for the output neuron) are asymptotically the {\it potential matching} estimates
\[
\lim_{N\rightarrow\infty}\hat{\theta}_{ELM,\mathcal{T}_N} = \arg\min_{\theta}\lim_{N\rightarrow\infty}C(\theta|\mathcal{T}_N) = \hat{\theta}   
\]
since the Markov chain will eventually converge to the target distribution. When truncated at a finite $N$, the estimated parameters are almost the empirical {\it potential matching} estimates except that the samples from the history of the Markov chain are not exactly (but quite close) from the target distribution.

\section{Adaptive learning}\label{sec:OLproof}

\begin{theorem}[Greville]\label{thm:greville}
If a matrix A, with $k$ columns, is denoted by $A_k$ and partitioned as $A_k=[A_{k-1}\;a_k]$, with $A_{k-1}$ a matrix having $k-1$ columns, then the Moore-Penrose generalized inverse of $A_k$ is
\[
A_k^{\dagger} =\begin{bmatrix}A_{k-1}^{\dagger}(I-a_kb_k^T)\\b_k^T\end{bmatrix}
\]
where 
\[
c_k = (I-A_{k-1}A_{k-1}^{\dagger})a_k, \quad  b_k = \left\{\begin{array}{ll}\frac{(A_{k-1}^{\dagger})^TA_{k-1}^{\dagger}a_k}{1+ \|A_{k-1}^\dagger a_k\|^2}, &c_k=0\\ \frac{c_k}{\|c_k\|^2}, & c_k\neq 0\end{array}\right.
\]
\end{theorem}

{\bf Proof of Propositon \ref{prop:online}}\\
To save computational cost, we only update the estimator. Suppose the current output matrix is $H_{k+1}=\begin{bmatrix}H_k\\h_{k+1}^T\end{bmatrix}$ and the target vector is $T_{k+1}=\begin{bmatrix}T_k\\t_{k+1}\end{bmatrix}$, then 
\begin{align*}
v_{k+1}^TH_{k+1}^T = T_{k+1}^T \Rightarrow v_{k+1}^T &= T_{k+1}^T(H_{k+1}^T)^{\dagger} \\
&= [T_k^T\; t_{k+1}] \left([H_k^T\; h_{k+1}]\right)^{\dagger}\\
&= [T_k^T\; t_{k+1}] \begin{bmatrix}(H_k^T)^{\dagger}(I-h_{k+1}b_{k+1}^T)\\b_{k+1}^T\end{bmatrix}\\
&= T_k^T\left(H_k^T\right)^{\dagger}(I-h_{k+1}b_{k+1}^T) + t_{k+1}b_{k+1}^T\\
&= v_k^T + (t_{k+1}-v_k^Th_{k+1})b_{k+1}^T
\end{align*}
where $b_{k+1}$ is obtained according to Theorem \ref{thm:greville}. Note that the computation of $b_{k+1}$ and $c_{k+1}$ still involves $H_k$ and $H_k^{\dagger}$ whose sizes increase as data size grows. Following \cite{kohonen88, kovanic79, vanschaik15}, we introduce two auxiliary matrices here
\[
\Phi_{k} = I-H_k^T(H_k^T)^{\dagger} \in \mathbb{R}^{s\times s},\quad \Theta_{k} = \left((H_k^T)^{\dagger}\right)^T(H_k^T)^{\dagger}= H_k^{\dagger}(H_k^T)^{\dagger} \in\mathbb{R}^{s\times s}
\]
and rewrite $b_{k+1}$ and $c_{k+1}$ as 
\[
c_{k+1} = \Phi_{k}h_{k+1}, \quad  b_{k+1} = \left\{\begin{array}{ll}\frac{\Theta_kh_{k+1}}{1+ h_{k+1}^T\Theta_kh_{k+1}}, &c_{k+1}=0\\ \frac{c_{k+1}}{\|c_{k+1}\|^2}, & c_{k+1}\neq 0\end{array}\right.
\]
Updating of the two auxiliary matrices can also be done adaptively
\[
\Phi_{k+1} = I - H_{k+1}^T(H_{k+1}^T)^{\dagger} = I-[H_k^T\;h_{k+1}]\begin{bmatrix}(H_k^T)^{\dagger}(I-h_{k+1}b_{k+1}^T)\\b_{k+1}^T\end{bmatrix} = \Phi_k -\Phi_kh_{k+1}b_{k+1}^T
\]
\[
\Theta_{k+1} = H_{k+1}^{\dagger}(H_{k+1}^T)^{\dagger} = [(I-b_{k+1}h_{k+1}^T)H_k^{\dagger}\;\;b_{k+1}]\begin{bmatrix}(H_k^T)^{\dagger}(I-h_{k+1}b_{k+1}^T)\\b_{k+1}^T\end{bmatrix} = (I-b_{k+1}h_{k+1}^T)\Theta_k(I-h_{k+1}b_{k+1}^T) + b_{k+1}b_{k+1}^T 
\]
and if $c_{k+1} = 0$, these formulas can be simplified as below
\[
\Phi_{k+1} = \Phi_k,\quad \Theta_{k+1} = \Theta_k - \Theta_kh_{k+1}b_{k+1}^T
\]

\end{document}